\documentclass[aps,prd,showpacs,floatfix,nofootinbib,12pt]{revtex4-1}

\usepackage{cmap}
\usepackage[T1]{fontenc}
\usepackage{graphicx}
\usepackage[caption=false]{subfig}
\usepackage{xcolor}  
\usepackage[latin1]{inputenc}
\usepackage{mathtools}  
\usepackage{hyperref}
\usepackage{epstopdf}
\usepackage{geometry}
\usepackage{ulem}
\usepackage{amssymb}

\usepackage{adjustbox}

\newcommand{\lsim}{\mathrel{\mathop{\kern 0pt \rlap
  {\raise.2ex\hbox{$<$}}}
  \lower.9ex\hbox{\kern-.190em $\sim$}}}
\newcommand{\gsim}{\mathrel{\mathop{\kern 0pt \rlap
  {\raise.2ex\hbox{$>$}}}
  \lower.9ex\hbox{\kern-.190em $\sim$}}}

\newcommand{\mev}{\ensuremath{\;\mathrm{MeV}}}
\newcommand{\gev}{\ensuremath{\;\mathrm{GeV}}}
\newcommand{\tev}{\ensuremath{\;\mathrm{TeV}}}

\definecolor{darkgreen}{HTML}{009900}

\def  \bcen   {\begin{center}}
\def  \ecen   {\end{center}}
\def  \beq    {\begin{equation}}
\def  \eeq    {\end{equation}}
\def  \bpm    {\begin{pmatrix}}
\def  \epm    {\end{pmatrix}}
\def  \beqa   {\begin{eqnarray}}
\def  \eeqa   {\end{eqnarray}}

\def\bea{\begin{eqnarray}}
\def\eea{\end{eqnarray}}

\def \stu {St\"{u}eckelberg }

\def\De {\Delta}

\def\to {\rightarrow}

\def\arccot {arccot}

\begin{document}


\title{
Consistency of Gauged Two Higgs Doublet Model: \\Gauge Sector}
\author{
Cheng-Tse Huang$^1$, Raymundo Ramos$^2$, \\
Van Que Tran$^{2,3}$, Yue-Lin Sming Tsai$^{2,4}$ and Tzu-Chiang Yuan$^2$
}

	\affiliation{
\small{
$^1$Interdisciplinary Program of Sciences, National Tsing Hua University, Hsinchu 30013, Taiwan\\}
{$^2$Institute of Physics, Academia Sinica, Nangang, Taipei 11529, Taiwan\\}
{$^3$School of Physics, Nanjing University, Nanjing 210093, China\\}
{$^4$Key Laboratory of Dark Matter and Space Astronomy, Purple Mountain Observatory, Chinese
Academy of Sciences, Nanjing 210008, China}
}

\date{\today}

\begin{abstract}
We study the constraints on the new parameters in the gauge sector of 
gauged two Higgs doublet model using the electroweak precision test data collected 
from the Large Electron Positron Collider (LEP) at and off the $Z$-pole 
as well as the current Drell-Yan and high-mass dilepton resonance 
data from the Large Hadron Collider (LHC).
Impacts on the new parameters by the projected sensitivities of various electroweak observables 
at the Circular Electron Positron Collider (CEPC) proposed to be built in China 
are also discussed. 
We also clarify why the St\"{u}eckelberg mass $M_Y$ for the hypercharge $U(1)_Y$ is set 
to be zero in the model by showing that it would otherwise lead to the violation of the 
standard charge assignments for the elementary quarks and leptons when they couple to the massless 
photon.

\end{abstract}

\maketitle

\section{Introduction} \label{section:intro}

The discovery of the 125 GeV scalar boson identified as the Higgs boson in the Standard Model (SM)
~\cite{Salam:1959zz,Glashow:1961tr,Weinberg:1967tq,Salam}
suggested that the simple Higgs mechanism~\cite{Englert:1964et,Higgs:1964pj,Guralnik:1964eu}
for electroweak symmetry breaking proposed by Weinberg~\cite{Weinberg:1967tq} and Salam~\cite{Salam} 
is the choice by {\it nature}.
Both Run I and Run II data collected by the two experimental groups ATLAS and CMS at the 
Large Hadron Collider (LHC) reveal no significant deviations from the SM predictions.
Alternative models for electroweak symmetry breaking like technicolor or composite Higgs models 
are arguably more elegant but necessarily more complicated. 
Simplicity seems to be more superior over other criterion like complexity or elegance for model buildings.

Nevertheless experimental observations of neutrino oscillations imply there must be new physics beyond the SM
to account for the minuscule masses of neutrinos. Missing mass problem and cosmic acceleration of our universe
also suggested the introduction of dark matter (DM)~\cite{Lin:2019uvt} and dark energy~\cite{Peebles:2002gy}. 
The standard $\Lambda$CDM model of cosmology~\cite{Dodelson:2003ft} 
consists of the SM of particle physics plus two new ingredients, namely the cold dark matter, 
which can be the weakly interacting massive particle predicted by many new particle physics models,
and a tiny positive cosmological constant at the present time in the Einstein's field equation for gravity,
which can be mimicked by numerous models of dark energy.
Many models of dark matter and neutrino masses require extension not only 
of the simple Higgs sector but sometimes also the electroweak gauge sector of the SM as well.
Moreover, models of dark energy are often represented by new scalar field with 
equation of state that can provide negative pressure in order to explain the cosmic acceleration at late times.

Thus extension of the SM in one way or the other seems necessary if one wants to solve the 
above puzzles in the neutrino sector and in cosmology.
At the same time, one should be open-minded that there might be other approaches other than particle physics
to answer some of these questions and remembering that  {\it nature} is the ultimate arbiter of all theoretical imaginations.

The gauged two Higgs doublet model (G2HDM) proposed in~\cite{Huang:2015wts} was motivated partly 
by the inert Higgs doublet model (IHDM)~\cite{Deshpande:1977rw,Ma:2006km,Barbieri:2006dq,LopezHonorez:2006gr} 
of dark matter. 
IHDM is a variant of the general 2HDM~\cite{Branco:2011iw} with an imposed discrete $Z_2$ symmetry on the scalar potential
and the Yukawa couplings such that one of the Higgs doublets is odd and become a scalar dark matter candidate. 
Dangerous tree level flavor changing neutral current (FCNC) interactions in the Yukawa couplings, generally presence in the general 2HDM,
are also eliminated by this discrete symmetry.
Due to its relatively simple extension of the SM, many detailed analysis of IHDM had been done in the literature
~\cite{Arhrib:2013ela,Arhrib:2014pva,Ilnicka:2015jba,Belyaev:2016lok,Eiteneuer:2017hoh,Borah:2018rca,Kephart:2015oaa,Goudelis:2013uca,Swiezewska:2012eh,Arhrib:2012ia}. 
In G2HDM, the discrete $Z_2$ symmetry in IHDM was not enforced. Instead the two Higgs doublets $H_1$ and $H_2$
are grouped into a two-dimensional irreducible representation $H = (H_1, \, H_2)^{\rm T}$ of a new gauge group $SU(2)_H$.
A priori there is no need to impose the discrete $Z_2$ symmetry in G2HDM. 
Once we write down all renormalizable interactions for G2HDM,
this discrete symmetry emerges as an accidental symmetry automatically.
Tree level flavor changing neutral current (FCNC) interaction in the Higgs-Yukawa couplings are also absence naturally for the SM fermions.
As long as one does not break this symmetry spontaneously,
which might lead to the domain wall problem in early universe,
the $H_2$ doublet is naturally an inert Higgs doublet and can play some role in dark matter physics. 
It is more satisfactory to have a global discrete symmetry like the $Z_2$ parity that guarantees the stability of dark matter
embedded into a local symmetry. Indeed there exists theoretical arguments showing that global continuous or discrete symmetries
are not compatible with quantum gravity~\cite{Krauss:1988zc,Kallosh:1995hi}.
Detailed analysis of the complex scalar dark matter physics in G2HDM will be presented in a forthcoming paper~\cite{DMPaper}.

The construction of G2HDM in~\cite{Huang:2015wts} involves extension of both the Higgs and gauge sector of the SM
which we will discuss shortly in the next section.
Several phenomenological implications of G2HDM had been explored in~\cite{Huang:2015rkj,Huang:2017bto,Arhrib:2018sbz,Chen:2018wjl}. 
In particular, we have studied recently in details the theoretical and phenomenological constraints
on the scalar sector~\cite{Arhrib:2018sbz}. 

We note that the 2HDM augmented with an extra local abelian $U(1)_X$ 
has been discussed in the literature~\cite{Ko:2012hd,Campos:2017dgc,Camargo:2018klg,Camargo:2018uzw,Camargo:2019ukv,Cogollo:2019mbd} to address neutrino masses, dark matter and to avoid FCNC interactions at the tree level.

As mentioned before, all experimental data are in line with SM predictions. 
The extended gauge sector of G2HDM must be challenged by electroweak precision test (EWPT) data
obtained previously at LEP-I and LEP-II as well as current data at the LHC. 
Constraints must be imposed on the new parameters in the extended gauge sector of G2HDM. 
The main purpose of this work is to study these constraints on the gauge sector 
systematically in analogous to previous analysis~\cite{Arhrib:2018sbz} done for the scalar sector. 
It is also interesting to address the sensitivities of these new parameters at the future colliders.
 
The contents of this paper is organized as follows: 
In the next Sec.~\ref{section:model}, we review the G2HDM and highlight some of its crucial features
of the gauge sector relevant most to this work.
Sec.~\ref{section:constraints} discusses the experimental constraints, 
including the electroweak precision test constraints at and off the $Z$-pole at LEP, 
Drell-Yan data from on-shell decay of the $Z$ boson at the LHC,
and the full  LHC Run II data from the high-mass dilepton resonance of an extra neutral gauge boson $Z^\prime$. 
Sec.~\ref{section:results} contains our numerical results from the profile likelihood analysis.
We also study future sensitivities of the new parameters in future experiments, 
in particular for the Circular Electron Positron Collider (CEPC)~\cite{CEPC-SPPCStudyGroup:2015csa} proposed/debated 
to be built in China.
Finally, we summarize and conclude in Sec.~\ref{section:conclusions}.
In Appendix~\ref{appendix:euler} we present
the formulas for the mixing angles among the three massive neutral gauge bosons in G2HDM
in terms of the fundamental parameters in the Lagrangian of the model.
The dominant two-body decay widths for the two new neutral gauge bosons 
are discussed in Appendix~\ref{appendix:decaywidths}.

\section{G2HDM Set Up}  \label{section:model}

In this section, we will start with a brief review for the set-up of G2HDM~\cite{Huang:2015wts} by specifying its particle content 
(Sec.~\ref{subsection:particlecontent}) and then write down the mass spectrum of the neutral gauge bosons 
(Sec.~\ref{subsection:masses}) and their interactions with the SM fermions (Sec.~\ref{subsection:neutralcurrents}) in the model.
Along the way, we will discuss some peculiar effects for nonzero \stu mass $M_Y$ associated with the hypercharge $U(1)_Y$. 

\begin{table}[t]
\begin{tabular}{|c|c|c|c|c|c|c|}
\hline
Fields & Spin & $SU(3)_C$ & $SU(2)_L$ & $SU(2)_H$ & $U(1)_Y$ & $U(1)_X$ \\
\hline\hline
$H=\left( H_1 \;\; H_2 \right)^{\rm T}$ & 0 & 1 & 2 & 2 & $\frac{1}{2}$ & $1$ \\
$\Delta_H=\left( \begin{array}{cc} \Delta_3/2 & \Delta_p/\sqrt{2}  \\ \Delta_m/\sqrt{2} & - \Delta_3/2 \end{array} \right)$ & 0 & 1 & 1 & 3 & 0 & 0 \\
$\Phi_H=\left( \Phi_1 \;\; \Phi_2 \right)^{\rm T}$ & 0 & 1 & 1 & 2 & 0 & $1$ \\
\hline \hline
$Q_L=\left( u_L \;\; d_L \right)^{\rm T}$ & $\frac{1}{2}$ & 3 & 2 & 1 & $\frac{1}{6}$ & 0\\
$U_R=\left( u_R \;\; u^H_R \right)^{\rm T}$ & $\frac{1}{2}$  & 3 & 1 & 2 & $\frac{2}{3}$ & $1$ \\
$D_R=\left( d^H_R \;\; d_R \right)^{\rm T}$ & $\frac{1}{2}$  & 3 & 1 & 2 & $-\frac{1}{3}$ & $-1$ \\
\hline
$u_L^H$ & $\frac{1}{2}$  & 3 & 1 & 1 & $\frac{2}{3}$ & 0 \\
$d_L^H$ & $\frac{1}{2}$  & 3 & 1 & 1 & $-\frac{1}{3}$ & 0 \\
\hline
$L_L=\left( \nu_L \;\; e_L \right)^{\rm T}$ & $\frac{1}{2}$  & 1 & 2 & 1 & $-\frac{1}{2}$ & 0 \\
$N_R=\left( \nu_R \;\; \nu^H_R \right)^{\rm T}$ & $\frac{1}{2}$  & 1 & 1 & 2 & 0 & $1$ \\
$E_R=\left( e^H_R \;\; e_R \right)^{\rm T}$ & $\frac{1}{2}$  & 1 & 1 & 2 &  $-1$  &  $-1$ \\
\hline
$\nu_L^H$ & $\frac{1}{2}$  & 1 & 1 & 1 & 0 & 0 \\
$e_L^H$ & $\frac{1}{2}$  & 1 & 1 & 1 & $-1$ & 0 \\
\hline\hline
$g^a_\mu (a=1, \cdots , 8) $ & 1 & 8 & 1 & 1 & 0 & 0 \\
$W^i_\mu (i=1,2,3)$ & 1 & 1 & 3 & 1 & 0 & 0 \\
$W^{\prime i}_\mu (i=1,2,3) $ & 1 & 1 & 1 & 3 & 0 & 0 \\
$B_\mu$ & 1 & 1 & 1 & 1 & 0 & 0 \\
$X_\mu$ & 1 & 1 & 1 & 1 & 0 & 0 \\
\hline
\end{tabular}
\caption{Particle content and their quantum number assignments in G2HDM. 
}
\label{tab:quantumnos}
\end{table}

\subsection{Particle Content \label{subsection:particlecontent}}

The particle content of G2HDM is listed in Table~\ref{tab:quantumnos}~\footnote{
$u^H_L,d^H_L,\nu^H_L,e^H_L$ in the table were denoted as $\chi_u,\chi_d,\chi_\nu,\chi_e$ respectively 
in~\cite{Huang:2015wts}.}.
Besides the two Higgs doublets $H_1$ and $H_2$ combining to form $H=(H_1,
H_2)^{\rm T}$ in the fundamental 
representation of an extra $SU(2)_H$, we introduced a triplet $\De_H$ and a doublet $\Phi_H$ of this new gauge group.
However $\De_H$ and $\Phi_H$ are singlets under the electroweak SM gauge group $SU(2)_L \times U(1)_Y$. 
Only $H$ carries both quantum numbers of the $SU(2)_L$ and $SU(2)_H$. 

There are different ways of introducing new heavy fermions in the model but we choose a
simple realization: the heavy fermions together with the SM right-handed fermions comprise $SU(2)_H$ doublets,
while the SM left-handed doublets are singlets under $SU(2)_H$. 
We note that heavy right-handed neutrinos paired up with 
a mirror charged leptons forming $SU(2)_L$ doublets was suggested before in the mirror fermion model~\cite{Hung:2006ap}.
To render the model anomaly-free, four additional chiral (left-handed) fermions for each generation, 
all singlets under both $SU(2)_L$ and $SU(2)_H$, are included.
For the Yukawa interactions that couple among the fermions and scalars in G2HDM, 
we refer our readers to~\cite{Huang:2015wts} for more details, since they are not relevant to this work.

To avoid some unwanted pieces in the scalar potential and Yukawa couplings, 
we  require the matter fields to  carry extra local $U(1)_X$ charges. Thus the complete 
gauge groups in G2HDM consist of $SU(3)_C \times SU(2)_L \times U(1)_Y  \times SU(2)_H \times U(1)_X$.
Apart from the matter content of G2HDM,
there also exist the gauge bosons corresponding to the SM and the extra 
gauge groups. 

The salient features of G2HDM are:
(i) it is free of gauge and gravitational anomalies; 
(ii) renormalizable;
(iii) without resorting to an ad-hoc $Z_2$ symmetry, 
an inert Higgs doublet $H_2$ can be naturally realized, providing a DM candidate;
(iv) due to the nonabelian $SU(2)_H \times U(1)_X$ gauge symmetry, 
dangerous FCNC interactions are absent at tree level for the SM sector;
(v) the VEV of the triplet can trigger $SU(2)_L$ symmetry breaking while that of $\Phi_H$ provides 
a mass to the new fermions through $SU(2)_H$-invariant Yukawa couplings;
{\it etc.}

\subsection{Neutral Gauge Boson Masses \label{subsection:masses}}

Consider the interaction basis $\{B, W^3, W^{\prime 3}, X\}$ for the 
neutral gauge bosons and denote their mass eigenstates as $\{A,Z_1,Z_2,Z_3\}$. 
After spontaneous symmetry breaking, the 4$\times$4 mass matrix 
in the interaction basis of $\{B, W^3, W^{\prime 3}, X\}$ is given by~\cite{Huang:2015wts}
\begin{equation}
\label{eq:M1sq2}
\mathcal{M}_\text{gauge}^2 =
\begin{pmatrix}
    \frac{g^{\prime 2} v^2 }{4} + M_Y^2 &
    - \frac{g^{\prime} g \, v^2 }{4}  &
    \frac{g^{\prime} g_H v^2 }{4} &
    \frac{g^\prime g_X v^2}{2} + M_X M_Y \\
    - \frac{g^{\prime} g \, v^2 }{4} &
    \frac{ g^2 v^2 }{4} &
    - \frac{g g_H v^2 }{4} &
    - \frac{ g g_X v^2  }{2} \\
    \frac{g^{\prime} g_H v^2 }{4} &
    - \frac{g g_H v^2 }{4} &
    \frac{g^2_H \left( v^2 + v^2_\Phi \right) }{4} &
    \frac{g_H g_X \left( v^2 - v^2_\Phi \right) }{2} \\ 
    \frac{g^\prime g_X v^2}{2} + M_X M_Y &
    - \frac{ g g_X v^2  }{2} &
    \frac{g_H g_X \left( v^2 - v^2_\Phi \right) }{2} &
    g_X^2 \left( v^2 + v^2_\Phi \right) + M_X^2
\end{pmatrix} \; .
\end{equation}
Here $g$, $g^\prime$, $g_H$ and $g_X$ denote the gauge couplings of
$SU(2)_L$, $U(1)_Y$, $SU(2)_H$ and $U(1)_X$ respectively;
$v$ and $v_\Phi$ are the vacuum expectation values (VEVs)
of $H_1$ and $\Phi_H$ respectively;
$M_X$ and $M_Y$ are the St\"{u}eckelberg masses for the two abelian $U(1)_X$ and $U(1)_Y$ respectively.
We note that $v_\Delta$ the VEV of the triplet $\Delta_H$ does not enter into the neutral gauge boson mass matrix.
This is unlike the case of scalar boson mass matrix analyzed in~\cite{Arhrib:2018sbz} 
which involves all three VEVs, $v$, $v_H$ and $v_\Delta$.
The matrix $\mathcal{M}^2_{\rm gauge}$ in Eq.~(\ref{eq:M1sq2}) is real and symmetric and thus 
can be diagonalized by a 4$\times$4 orthogonal rotation matrix that we will
denote as ${\cal O}^{4\times 4}$
\begin{equation}
    \label{eq:diagmgauge}
    ({\cal O}^{4\times 4})^{\rm T} \cdot \mathcal{M}^2_\text{gauge} \cdot {\cal O}^{4\times 4} = \text{diag}(0, M^2_{Z_1}, M^2_{Z_2}, M^2_{Z_3}) \, ,
\end{equation}
where $M^2_{Z_1} <M^2_{Z_2} < M^2_{Z_3}$. The zero mass state is naturally identified as the photon.

Some comments on the St\"{u}eckelberg masses $M_X$ and $M_Y$ are in order here.
It has been demonstrated in~\cite{Ruegg:2003ps} that for the extension of SM with
a St\"{u}eckelberg mass $M_Y$ for the hypercharge $U(1)_Y$, 
there exists a plethora of new physical effects. 
Notably, besides the photon obtaining a mass, neutrinos will couple to the photon and 
charged leptons will have axial vector couplings with the photon. 
Nevertheless, the St\"{u}eckelberg extension of the SM doesn't spoilt renormalizability of the model.
All these new effects are proportional to $M_Y$. Experimentally, the photon mass upper bound 
deduced from modeling the solar wind in magnetohydrodynamics
is $m_\gamma < 1 \times 10^{-18}$ eV~\cite{Tanabashi:2018oca},
which implies $M_Y$ must be very tiny too.  
If individual St\"{u}eckelberg mechanism is introduced for each of the two $U(1)$s factors in G2HDM, 
the photon will in general obtain nonzero mass and many results obtained in~\cite{Ruegg:2003ps} apply as well.
In~\cite{Huang:2015wts}, we followed~\cite{Kors:2005uz,Kors:2004iz,Kors:2004ri,Kors:2004dx} in which 
only one St\"{u}eckelberg field was introduced for the two factors of $U(1)$s to implement the St\"{u}eckelberg mechanism.
The matrix $\mathcal{M}_\text{gauge}^2$ thus obtained given in Eq.~(\ref{eq:M1sq2}) has zero determinant 
and a massless photon can always be realized for arbitrary values of the St\"{u}eckelberg masses $M_X$ and $M_Y$.

In the next subsection, we will show that with a nonzero $M_Y$ the electric charge assignments of the
SM fermions and their heavy partners in G2HDM will no longer be standard but instead receive milli-charge
corrections like those discussed in~\cite{Ruegg:2003ps}.
In particular, neutrinos will couple to the photon and all fermions also have axial vector couplings with the photon at tree level.
These peculiar effects depend on $M_Y$ through the mixing matrix elements and hence necessarily small. 
Thus, we have strong theoretical motivation to set $M_Y = 0$ in what follows to avoid these unpleasant features.
For an analysis with both $M_X$ and $M_Y$ nonzero in a St\"{u}eckelberg $U(1)_X$ extension of the SM 
that maintains the standard QED interaction for the SM fermions,
see~\cite{Feldman:2007nf,Feldman:2007wj,Feldman:2006wb}.
The main reason why the photon-fermion couplings in 
G2HDM are in general different from these previous works is due to the presence of the extra gauge group $SU(2)_H$
whereas there is only one extra abelian group $U(1)_X$ in ~\cite{Feldman:2007nf,Feldman:2007wj,Feldman:2006wb}.

Setting $M_Y = 0$ in G2HDM will simplify $\mathcal{M}_\text{gauge}^2$ and allows us 
to write the rotation matrix in the following product form
\begin{equation}
\label{eq:rot4by4dec}
{\cal O}^{4\times 4}_{M_Y=0} =
\begin{pmatrix}
    c_W & -s_W & 0 & 0 \\
    s_W & c_W & 0 & 0 \\
    0 & 0 & 1 & 0 \\
    0 & 0 & 0 & 1
\end{pmatrix}
\cdot
\begin{pmatrix}
    1 & 0 & 0 & 0 \\
    0 &  &  &\\
    0 &  &  {\cal O} &\\
    0 &  &  &
\end{pmatrix},
\end{equation}
where $c_W$ and $s_W$ represent $\cos\theta_W$ and $\sin\theta_W$
respectively, with $\theta_W$ being the Weinberg angle defined by
\begin{equation}
\label{weinbergangle}
e^{\rm SM} \equiv g \sin\theta_W = g^\prime \cos\theta_W = \frac{g g^\prime}{\sqrt {g^2 + g^{\prime 2}} } \; .
\end{equation}
It is obvious that the matrix ${\cal O}^{4\times 4}_{M_Y=0}$ in Eq.~(\ref{eq:rot4by4dec}) 
is just the product of the SM gauge rotation matrix 
made into a $4\times4$ matrix, 
called ${\cal O}^{4\times 4}_\text{SM}$, times a general $3\times 3$
orthogonal rotation matrix ${\cal O}$ which was also converted to a
$4\times 4$ matrix. After applying the rotation ${\cal O}^{4\times
4}_\text{SM}$ to $\mathcal{M}^2_{\text{gauge}}(M_Y=0)$, the result is
\begin{align}
    {\cal O}^{4\times 4 \, \rm T}_\text{SM} \cdot \mathcal{M}^2_\text{gauge} (M_Y=0)  \cdot {\cal O}^{4\times 4}_\text{SM} & =
\begin{pmatrix}
    0 & 0 & 0 & 0\\
    0 &
    M_{Z^\text{SM}}^2 &
        - \frac{g_H v }{2} M_{Z^\text{SM}} &
        - g_X v M_{Z^\text{SM}} \\
    0 &
        - \frac{g_H v}{2} M_{Z^\text{SM}} &
        \frac{g_H^{2} \left(v^{2} + v_\Phi^{2}\right)}{4} &
        \frac{g_X g_H \left(v^{2} - v_\Phi^{2}\right)}{2}\\
    0 &
        - g_X v M_{Z^\text{SM}} &
        \frac{g_X g_H \left(v^{2} - v_\Phi^{2}\right)}{2} &
        g_X^{2} (v^{2} + v_\Phi^{2}) + M_X^{2}
\end{pmatrix} \, ,
\label{eq:MgaugeSMrot}
\end{align}
where $M_{Z^\text{SM}} = v\sqrt{g^2+g^{\prime 2}} /2$ is the mass of the $Z$ boson in the SM. We can consider
the vanishing (1,1) element to be the mass of the photon eigenstate $A_\mu$.
Furthermore, according to Eqs.~\eqref{eq:diagmgauge} and
\eqref{eq:rot4by4dec}, the remaining 3$\times$3 matrix formed by the
non-vanishing elements above is diagonalized by the orthogonal matrix $ {\cal O}$. 
In particular, one can parametrize $ {\cal O}$ in terms of the following Tait-Bryan representation
\begin{equation}
\label{eq:Rpara}
 {\cal O}=
\begin{pmatrix}
    c_{\psi} c_{\phi}- s_{\theta} s_{\phi} s_{\psi}& -c_{\theta} s_{\phi}  &  s_{\psi} c_{\phi}+ s_{\theta} s_{\phi} c_{\psi}\\
    c_{\psi} s_{\phi}+ s_{\theta} c_{\phi} s_{\psi}  &c_{\theta} c_{\phi} &  s_{\psi} s_{\phi}- s_{\theta} c_{\phi} c_{\psi}\\
    -c_{\theta} s_{\psi} & s_{\theta} &  c_{\theta} c_{\psi}
\end{pmatrix},
\end{equation}  
where $s_x$ and $c_x$ stand for sine and cosine with the rotation angle $x=\phi,\theta,\psi$ respectively. 
As shown in Appendix~\ref{appendix:euler}, these rotation angles can be represented as
\beq
\label{eq:Tanphi}
\tan({\phi}) = \frac{- g_H v M_{Z^\text{SM}} (M_X^2 - M_{Z_2}^2 + 2 g_X^2 v_\Phi^2)}
{2 \Big( M_{Z_2}^4 -  \big( M_{Z^\text{SM}}^2 + M_X^2 + (v^2+v_\Phi^2) g_X^2 \big) M_{Z_2}^2 +  M_{Z^\text{SM}}^2 (M_X^2 + g_X^2 v_\Phi^2)\Big)}\, ,
\end{equation} 
\begin{equation}
\label{eq:Tantheta}
\tan({\theta}) = \frac{-g_X (M_{Z_2}^2 (v^2-v_{\Phi}^2) + M_{Z^\text{SM}}^2 v_{\Phi}^2)}
{v M_{Z^\text{SM}} (M_X^2 - M_{Z_2}^2 + 2 g_X^2 v_\Phi^2)} \sin{\phi} \, ,
\end{equation}
\begin{equation}
\label{eq:Cotpsi}
\cot({\psi}) =  \frac{g_H(M_{Z_1}^2-M_X^2 - 2 g_X^2 v_\Phi^2)}{g_X ( g_H^2 v_\Phi^2 - 2 M_{Z_1}^2)} \frac{\cos{\theta}}{\sin{\phi}} 
- \sin{\theta} \cot{\phi} \, .
\end{equation} 


It is easy to see that taking the limits of $g_H$ and $g_X$ go to 0, the non-vanishing 3$\times$3
block matrix in Eq.~\eqref{eq:MgaugeSMrot} becomes ${\rm Diag}(M_{Z^{\rm SM}}^2, 0, M_X^2)$.
Thus the rotation matrix $\mathcal O$ must be identity. This can be realized by setting 
$\phi$, $\theta$ and $\psi$ to be zeros which can be derived from  
Eqs.~\eqref{eq:Tanphi}, \eqref{eq:Tantheta} and \eqref{eq:Cotpsi}.

We note that if one sets $M_X$ to zero, the mass matrix in the right-handed side of Eq.~\eqref{eq:MgaugeSMrot} 
is symmetric under the interchange of $g_H/2 \leftrightarrow g_X$.

After the rotation matrix ${\cal O}$ is found, the $Z_i$ mass eigenstates where $i$ runs from 1 to 3 are given by 
\beq
(Z_1,Z_2,Z_3)^{\rm T} = {\cal O}^{\rm T} \cdot (Z^{\rm SM}, W^{\prime 3}, X)^{\rm T} \, .
\label{eq:Zcomposition}
\eeq
The composition $Z^{\rm SM}$, $W^{\prime 3}$ and $X$ of the $Z_i$ mass eigenstate
is given by ${\cal O}_{1i}^2$, ${\cal O}_{2i}^2$, and ${\cal O}_{3i}^2$, respectively. 
In general, the $Z$-pole can be any one of the $Z_i$ depending on which one is actually 
closer to the pole by the underlying parameter choices in G2HDM. 
In our analysis, we will consider there is always at least one extra neutral gauge boson heavier than the $Z$-pole.


\subsection{Neutral Gauge Current Interactions \label{subsection:neutralcurrents}}

The part of the Lagrangian that contains the interaction of the $Z_i$ with visible matter in G2HDM is
\begin{equation}
    \mathcal{L}_{N} = g_M \sum_{f}\sum^3_{i=1}\bar{f}\gamma_\mu \left[
        \left(v_{f}^{(i)} - \gamma_5 a^{(i)}_f\right)Z_{i}^\mu\right]f \, ,
\label{eq:lagzff}
\end{equation}
where $g_M=\sqrt{g^2
+ g'^2}/2$. The $v^{(i)}_f$ and $a^{(i)}_f$ factors are given by ($M_Y \neq 0$)
\begin{align}
    \label{eq:vecfac}
    v^{(i)}_f = {}&
    \left(c_W {\cal O}^{4\times 4}_{2,i+1} - s_W {\cal O}^{4\times 4}_{1,i+1}\right) T^3_f
    + 2 Q_fs_W {\cal O}^{4\times 4}_{1,i+1}\nonumber\\
    & + \frac{1}{\sqrt{g^2 + g^{\prime 2}}}\left(
                            X_R g_X {\cal O}^{4\times 4}_{4,i+1}
                            + T^{3H}_{f_R} g_H {\cal O}^{4\times 4}_{3,i+1}
                        \right),\\
    \label{eq:axifac}
    a^{(i)}_f = {}&
    \left(c_W {\cal O}^{4\times 4}_{2,i+1} - s_W {\cal O}^{4\times 4}_{1,i+1}\right) T^3_f \nonumber\\
    & - \frac{1}{\sqrt{g^2 + g^{\prime 2}}}\left(
                            X_R g_X {\cal O}^{4\times 4}_{4,i+1}
                            + T^{3H}_{f_R} g_H {\cal O}^{4\times 4}_{3,i+1}
                        \right).
\end{align}
Here $T^3_f$ is the $SU(2)_L$ isospin charge and $Q_f$ is the electric charge
in units of $e^{\rm SM}$ for the SM fermion $f$ where 
$e^{\rm SM}$ is given by Eq.~(\ref{weinbergangle}). They are related to the $U(1)_Y$ hypercharge by the standard formula 
$Q^{\rm SM}_f = T^3_f + Y_f$. The charges due to the new gauge symmetries are $X_R$ as the $U(1)_X$
charge of the corresponding $f_R$ and $T^{3H}_{f_R}$ is the $SU(2)_H$
analogous of the $SU(2)_L$ isospin $T^3$ again for the corresponding $f_R$. 
We simply define $T^{3H}_{f_R}= \pm 1/2$ depending on $f_R$ belongs to
the upper or lower component of an $SU(2)_H$ doublet.

For the photon-fermion couplings in G2HDM, we obtain
\beq
{\cal L}_\gamma = - e^\text{SM} \sum_f \bar f \gamma_\mu \left( Q^{\rm G2HDM}_f - a^\gamma_f \gamma_5 \right) A^\mu f \; ,
\eeq
where
\begin{align}
\label{eq:photvecfac3}
Q^{\rm G2HDM}_f = {}&
\frac{{\cal O}^{4\times 4}_{1,1}}{c_W} Q^{\rm SM}_f + \frac{T^3_f}{2}\left(
	\frac{{\cal O}^{4\times 4}_{2,1}}{s_W}
	- \frac{{\cal O}^{4\times 4}_{1,1}}{c_W}\right) \nonumber \\
	& + \frac{1}{2 e^\text{SM}}\left(g_X  {\cal O}^{4\times 4}_{4,1} X_R 
	+ g_H {\cal O}^{4\times 4}_{3,1} T^{3H}_{f_R} \right) \; , \\
\label{eq:photaxifac3}
a^{\gamma}_f = {}&
\frac{T^3_f}{2}\left(\frac{{\cal O}^{4\times 4}_{2,1}}{s_W} - \frac{{\cal O}^{4\times
	4}_{1,1}}{c_W}\right) \nonumber \\
	& - \frac{1}{2 e^\text{SM}}\left(g_X X_R {\cal O}^{4\times 4}_{4,1}
		+ g_H T^{3H}_{f_R}{\cal O}^{4\times 4}_{3,1}\right) \; .
\end{align}
Thus, with both nonzero $M_X$ and $M_Y$, the electromagnetism interaction in G2HDM 
is in general different from the SM case.
The standard charge assignment for every SM fermion will  suffer from an overall 
correction factor of $\mathcal O^{4 \times 4}_{1,1}/c_W$ plus two correction terms, and there is also a non-vanishing
axial vector coupling.

Next, we can take the limit $M_Y = 0$ and write the corresponding
expressions. 
By replacing the elements of ${\cal O}^{4\times 4}$ by ${\cal O}^{4\times
4}_{1,1} = {\cal O}^{4\times 4}_{2,2} = c_W$ and $-{\cal O}^{4\times 4}_{1,2}
= {\cal O}^{4\times 4}_{2,1} = s_W$ as in Eq.~\eqref{eq:rot4by4dec}, one can
find the following new expressions for the vector and axial vector couplings
\begin{align}
    \label{eq:vecfacmy0}
    v^{(i)}_{f(M_Y = 0)} = {}&
     (T^3_f - 2 Q_fs_W^2){\cal O}_{1i}
     + \frac{1}{\sqrt{g^2 + g^{\prime 2}}}\left(
                            X_R g_X {\cal O}_{3i}
                            + T^{3H}_{f_R} g_H {\cal O}_{2i}
                        \right),\\
    \label{eq:axifacmy0}
    a^{(i)}_{f(M_Y = 0)} = {}&
    T^3_f  \,{\cal O}_{1i}
     - \frac{1}{\sqrt{g^2 + g^{\prime 2}}}\left(
                            X_R g_X {\cal O}_{3i}
                            + T^{3H}_{f_R} g_H {\cal O}_{2i}
                        \right).
\end{align}
Similarly, one can do the same substitutions on Eqs.~\eqref{eq:photvecfac3} and
\eqref{eq:photaxifac3} together with ${\cal O}^{4\times 4}_{3,1} = {\cal
O}^{4\times 4}_{4,1} = 0$ and check that the photon coupling to the SM fermions
goes back to the SM expression $Q^{\rm G2HDM}_f = Q^{\rm SM}_f = T^3_f + Y_f$ 
while all the axial vector couplings $a^\gamma_f$ vanish. This is the main physical reason 
why we set $M_Y = 0$ so as to reproduce the standard photon-fermion couplings.
For $M_X$, it can be arbitrary and is naturally to consider the light and heavy scenarios 
where it is smaller and greater than the $Z$-boson mass respectively.

Obviously, the formulas obtained in this subsection 
for the couplings of the neutral gauge bosons with 
the SM fermions also hold for the heavy fermions in G2HDM.

\section{The Constraints} \label{section:constraints}
\subsection{Constraints from Precision Electroweak Data \label{subsection:ewptdata} at LEP-I}
The interaction of $Z$ boson with SM fermions is described by the Lagrangian in
Eq.~\eqref{eq:lagzff}. For the case of $M_Y =0$ limit, the tree-level
couplings are shown in Eqs.~\eqref{eq:vecfacmy0} and \eqref{eq:axifacmy0}. For more precise calculation,
we include the radiation corrections from propagator self-energies and flavor
specific vertex corrections to the $Z$ boson and fermions couplings
\cite{Erler:2004nh,ALEPH:2005ab}, which now are given by~\footnote{We ignore loop corrections
related to the new gauge couplings $g_H$ and $g_X$.}
(suppressing $M_Y=0$ in the subscripts)
\begin{align}
    \label{eq:vzf}
    v_{f}^{i} = {}&
     \sqrt{\rho_{f}} \, (T^3_f - 2  \kappa_f Q_fs_W^2) {\cal O}_{1i}
     + \frac{1}{\sqrt{g^2 + g^{\prime 2}}}\left(
                            X_R g_X {\cal O}_{3i}
                            + T^{3H}_{f_R} g_H {\cal O}_{2i}
                        \right) \, ,\\
    \label{eq:azf}
    a_{f}^i = {}&
    \sqrt{\rho_{f}} \, T^3_f  \,{\cal O}_{1i}
     - \frac{1}{\sqrt{g^2 + g^{\prime 2}}}\left(
                            X_R g_X {\cal O}_{3i}
                            + T^{3H}_{f_R} g_H {\cal O}_{2i}
                        \right) \, ,
\end{align}
where $i$ in this work is either equal to 1 or 2 depending which
mass eigenstate is closest to $Z$-pole. The parameters $\rho_{f}$ and $\kappa_f$
are loop corrections quantities.
The decay of the $Z$ boson into fermions and anti-fermions in the on-shell
renormalization scheme is given by \cite{Baur:2001ze,Erler:2004nh}
\beq
 \Gamma(Z\to f\bar f) = N_f^c  \Gamma _o  \mathcal{R}_f  \sqrt {1-4\mu _f^2 } 
 \Bigg  [ |v_f|^2 (1 + 2\mu _f^2 ) + |a_f|^2 (1 -4\mu _f^2 )\Bigg] \, , 
 \label{eq:ZwidthRC}
 \eeq
where $N_f^c$ is the color factor (1 for leptons and 3 for quarks), 
$\Gamma_o = G_F M^3_Z/ 6\sqrt 2 \pi$, $\mu _f = m_f  / M_Z$ and 
\beq
 \mathcal{R}_f  = \left(1 + \delta _f^{QED} \right)\left(1 +\frac{N_f^c-1}{2}\delta _f^{QCD} \right) \, ,
 \eeq
 with 
 \bea
 \delta _{f}^{QED}  &=&\frac{3\alpha}{4\pi}Q_f^2 \, , \\
 \delta _{f}^{QCD}  &=& \frac{{\alpha_s }}{\pi } + 1.409\left( {\frac{{\alpha _s }}{\pi }} \right)^2
 -12.77\left( {\frac{{\alpha _s }}{\pi }} \right)^3  - Q_f^2
\frac{{\alpha \alpha _s   }}{{4\pi ^2 }} \, .
 \eea
Here $Q_f$ is the electric charge of the fermion $f$ in unit of $e^{\rm SM}$, and 
$\alpha$ and $\alpha _s $ are the fine-structure and strong coupling constants, 
respectively, evaluated at the $M_Z$ scale. 
It is understood that the couplings $v_f$ and $a_f$ in Eq.~(\ref{eq:ZwidthRC}) 
should be replaced by $v^i_f$ and $a^i_f$ in Eqs.~(\ref{eq:vzf}) and (\ref{eq:azf}) respectively 
with $i=1$ or 2 depending which $M_{Z_i}$ is closest to the $Z$-pole $M_Z$.

We also investigate some $Z$-pole ($\sqrt{s} \approx M_Z$) observables, including 
the ratio of partial decay width of $Z$ boson
\beq
R_{l} = \frac{\Gamma_{\rm had}}{\Gamma_{l^+l^-}} \, ,
\hspace{1cm}   
R_{q} = \frac{\Gamma_{q\bar q}}{\Gamma_{\rm had}} \, ,
\eeq
the hadronic cross-section
\beq
\sigma_{\rm had}=  \frac{12\pi \Gamma_{e^+e^-}\Gamma_{\rm had}}{M_Z^2 \, \Gamma_Z^2} \, ,
\eeq
the parity violation quantity
\beq 
A_{f}  =   \frac{2 v_f a_f}{v^2_f+a^2_f} \, ,
\eeq
and the forward-backward asymmetry quantity
\beq 
A_{\rm FB}  =   \frac{3}{4} A_f \frac{A_e + P_e}{1 + P_e A_e} \, , 
\eeq
where $P_e$ is the initial $e^-$ polarization. Recall that at LEP-I 
$P_e = 0$, {in this case} 
\beq
A^{(0,f)}_{\rm FB}  =\frac{3}{4} A_e A_f \, . 
\eeq
A summary of the electroweak observables at $Z$-pole from various experiments~\cite{Tanabashi:2018oca} 
is presented in Table \ref{table:ewobs}.

\begin{table}[th!]
	\centering		
	\begin{tabular}{ c | c | c | c  }
		\hline
		\hline
		Observables             & LEP Data     & CEPC Precision~\cite{CEPC-SPPCStudyGroup:2015csa} & Standard Model \\ 
		\hline
		$M_Z$ [GeV]       & 91.1876 $\pm$ 0.0021   &  $5\times 10^{-4}$  &  91.1884 $\pm$ 0.0020\\
		$\Gamma_Z$ [GeV]       & 2.4952 $\pm$ 0.0023 &  $5.06\times 10^{-4}$ & 2.4942 $\pm$ 0.0008 \\
		$ \Gamma_{\rm had}$ [GeV]  & 1.7444 $\pm$ 0.0020  &   --- & 1.7411 $\pm$ 0.0008\\
		$ \Gamma_ {\rm inv}$  [MeV]  & 499.0 $\pm$ 1.5     & ---  & 501.44 $\pm$ 0.04\\
		$ \Gamma_ {l^+l^-} $ [MeV] & 83.984 $\pm$ 0.086 &  --- & 83.959 $\pm$ 0.008\\
		$ \sigma_ {\rm had}[{\rm nb}] $      & 41.541 $\pm$ 0.037  & --- & 41.481 $\pm$ 0.008\\
		$R_e        $        & 20.804 $\pm$ 0.050  &  --- & 20.737 $\pm$ 0.010\\
		$R_\mu     $           & 20.785 $\pm$ 0.033  &  $0.05\%$  & 20.737 $\pm$ 0.010\\
		$R_\tau   $             & 20.764 $\pm$  0.045 &  $0.05\%$ & 20.782 $\pm$ 0.010 \\
		$R_b     $           & 0.21629 $\pm$ 0.00066   &  $0.08\%$ & 0.21582 $\pm$ 0.00002\\
		$R_c    $            & 0.1721 $\pm$ 0.0030   &  --- & 0.17221 $\pm$ 0.00003\\				
		$A^{(0,e)}_{\rm FB}    $      & 0.0145 $\pm$ 0.0025  &  --- & 0.01618 $\pm$ 0.00006\\
		$A^{(0,\mu)}_{\rm FB}  $        & 0.0169 $\pm$ 0.0013  & --- & 0.01618 $\pm$ 0.00006\\
		$A^{(0,\tau)}_{\rm FB}$          & 0.0188 $\pm$ 0.0017   &  --- & 0.01618 $\pm$ 0.00006\\
		$A^{(0,b)}_{\rm FB} $         & 0.0992 $\pm$ 0.0016   &   $0.15\%$ & 0.1030 $\pm$ 0.0002\\
		$A^{(0,c)}_{\rm FB} $        & 0.0707 $\pm$ 0.0035    &  --- & 0.0735 $\pm$ 0.0001\\
		$A^{(0,s)}_{\rm FB}$          & 0.0976 $\pm$ 0.0114   &  --- & 0.1031 $\pm$ 0.0002\\
		$A_e   $             & 0.15138 $\pm$ 0.00216  &    --- & 0.1469 $\pm$ 0.0003\\
		$A_\mu $               & 0.142 $\pm$ 0.015   &  --- & 0.1469 $\pm$ 0.0003\\
		$A_\tau$                & 0.136 $\pm$ 0.015  &    --- & 0.1469 $\pm$ 0.0003\\
		$A_b $               & 0.923 $\pm$ 0.020  &   --- & 0.9347\\
		$A_c $               & 0.670 $\pm$ 0.027  &  --- & 0.6677 $\pm$ 0.0001\\
		$A_s$                & 0.0895 $\pm$ 0.091 & --- & 0.9356\\ 
		\hline
		\hline      
	\end{tabular}	
	\caption {The electroweak observables at the $Z$-pole. 
	The second, third and last column are the LEP measurement~\cite{Tanabashi:2018oca}, 
	CEPC preliminary conceptual design report~\cite{CEPC-SPPCStudyGroup:2015csa}, 
	and the SM prediction~\cite{Tanabashi:2018oca}, respectively.
	}
	\label{table:ewobs}
\end{table}

From the data in Table~\ref{table:ewobs}, we build the
Chi-squared for the electroweak observables at $Z$-pole as follows
\bea
\chi^2_{Z-{\rm pole} }&=& \chi^2_{M_Z} + \chi^2_{\sigma_{\rm had}} + \max \left[ \chi^2_{\Gamma_Z}, (\chi^2_{\Gamma_{\rm had}}+\chi^2_{\Gamma_{\rm inv}}+\chi^2_{\Gamma_{l^+l^-}})\right]  \nonumber \\
&& +\sum_{f = (e,\mu,\tau,b,c)}{\chi^2_{R_f}}+ \sum_{f = (e,\mu,\tau,b,c,s)}{(\chi^2_{A_f} + \chi^2_{A_{\rm FB}^{(0,f)}})} \; .
\label{eq:chisqewobs}
\eea
Note that we have considered the correlations between the total decay width of $Z$ boson and its partial decay widths to 
hadrons, invisibles and dilepton. 
For each $\chi^2_i$ on the right-handed side of Eq.~(\ref{eq:chisqewobs}), it is given by 
the standard expression, namely
\bea
\chi^2_i = \frac{\left({O}_i^{\rm exp} - O_i^{\rm th}\right)^2}
{\left( \Delta O_i^{\rm exp} \right)^2} \; ,
\eea
where $O_i^{\rm exp/th}$ represents the experimental/theoretical value of 
any one of the 23  electroweak observables listed in Table~\ref{table:ewobs} and $\Delta O^{\rm exp}_i$ 
is the corresponding experimental uncertainty.

\subsection{Contact Interactions at LEP-II}

We also include constraints from data above the $Z$-pole by considering the
LEP-II measurements related to contact interactions taking 
the following form of effective Lagrangian
\begin{equation}
\label{eq:lagcont}
\mathcal{L}_{\mathrm{eff}} = \frac{\pm 4\pi}{
    (1 + \delta_{ef})(\Lambda^{\pm f}_{\alpha\beta})^2} \left(
    \bar{e} \gamma^\mu P_\alpha e \bar{f} \gamma_\mu P_\beta f
\right) \; ,
\end{equation}
where $P_{\alpha , \beta}$ represent the chirality projection operators 
with $\alpha,\beta$ being $L$ or $R$ for left-handed or right-handed fermions, respectively. 
The sign of Eq.~\eqref{eq:lagcont} depends on whether the interference between the contact
interaction it parametrizes and the SM process is constructive ($+$) or
destructive ($-$).
There is a total of 6 combinations for the $\alpha\beta$ indices of $\Lambda^{\pm
f}_{\alpha\beta}$: $\alpha\beta= \{{LL,LR,RL,RR,VV,AA} \}$, which are also called models.
The limits on $\Lambda^{\pm f}_{\alpha\beta}$ set by LEP-II are given in Table 3.15 of
Ref.~\cite{Schael:2013ita}. The strongest constraint is given by
$\Lambda^{+l}_{VV} > 24.6$ TeV.
By using these $\Lambda^{\pm f}_{\alpha\beta}$ values, we are able to
reconstruct the cross section for new physics processes based on the
Lagrangian in Eq.~\eqref{eq:lagcont}.

To improve the analysis of this section, in particular for the cases where the
mass of one of the gauge bosons is below the $Z$-pole, we calculate the
additional $Z$-like mediator contribution~\footnote{
In what follows, we will denote the extra neutral gauge boson as $Z^\prime$ or $Z_i$ 
depending on whether we refer to the experimental data or G2HDM.}
to the $e^-e^+\to Z_i \to f\bar{f}$ scattering
cross section. In the case $f=e$ we have the contribution of both $s$ and $t$
channels while for $f\neq e$ only the $s$ channel contributes.
Note that here we do not need the SM contributions such as the photon and $Z$
exchange not considered in Eq.~\eqref{eq:lagcont}.
In the massless approximation for all the external fermions, the amplitudes for the $s$ and $t$ channels and for the
interference term between them are given by: 
\begin{align}
\label{eq:Amps}
|\mathcal{M}_s|^2 & = \frac{2 g_M^4 \left\{\left[(a_f^{i})^4 + (v^i_f)^4\right]
   \left(s^2+2 s t+2 t^2\right)-2
   (a^i_f v^i_f)^2 \left(s^2+2 s
   t-2 t^2\right)\right\}}{\left(M_{Z_i}^2
   -s\right)^2} \, , \\
\label{eq:Ampt}
|\mathcal{M}_t|^2&  = \frac{2 g_M^4 \left[(a^i_f)^4
		\left(s^2+t^2\right)
		-2 (a^i_f v^i_f)^2 \left(s^2-3 t^2\right)
		+(v^i_f)^4
   \left(s^2+t^2\right)\right]}{\left(M_{Z_i}^2+s+t\right)^2}\, , \\
\label{eq:Ampst}
|\mathcal{M}_{st}|^2&  = \frac{4 g_M^4 t^2
	\left[(a^i_f)^4+6 (a^i_f v^i_f)^2+(v^i_f)^4\right]}{\left
   (M_{Z_i}^2-s\right)
   \left(M_{Z_i}^2+s+t\right)} \, ,
\end{align}
where $s$ is the center of mass energy squared, $t = s(\cos\varphi- 1)/2$ and
$\varphi$ is the angle between incoming and outgoing particles. This angle $\varphi$
should be integrated to obtain the final cross section. The resulting cross
section has to be compared against the cross section obtained using the effective
Lagrangian in Eq.~\eqref{eq:lagcont} with the $\Lambda^{\pm f}_{\alpha\beta}$
given by the experimental result. 
The couplings
$v^i_f$ and $a^i_f$ have
$i=$ 1 or 2 depending on whether we are analyzing light or heavy $M_X$
scenario. For $i=3$, we assume $M_{Z_3}$ is much heavier than $M_Z$ so that
its contributions are negligible. 
To be able to construct a $\chi^2$ from the LEP-II 95\% C.L. limit, we calculate
the corresponding 95\% C.L. cross section and compare against the theoretical
result. When our theoretical result matches the 95\% C.L. with null-signal
assumption, the corresponding
$\chi^2$ value should be 2.71~\footnote{
For a Gaussian distribution, the value of $\Delta\chi^2=2.71$ corresponds to the
90\% C.L. of a two-tailed test, but it also equivalent to the 95\% C.L. of a
one-tailed test that we are using.
}.
In this case, we calculate the $\chi^2$ value using
\begin{equation}
\chi^2_{\rm LEP-II} = 2.71 \times\left[ \frac{\sigma_{\text{G2HDM}}\left(e^+ e^-\to Z_i \to
f \bar f\right)}
{ \overline{\sum} \sigma_{\rm eff}\left(\Lambda^{\pm
f\text{(95\%)}}_{\alpha\beta}\right)}\right]^2 \; ,
\label{eq:chisqcontact}
\end{equation} 
where $\sigma_\text{eff}$ is the cross section obtained using the effective
Lagrangian of Eq.~\eqref{eq:lagcont} with the experimental results for
$\Lambda^{\pm f}_{\alpha\beta}$ given in Ref.~\cite{Schael:2013ita} for different
combinations of the chirality. The effective cross sections for different combinations of
$\alpha\beta=\{LL,RR,LR,RL\}$ from the data are summed and averaged. 
We do not consider the combinations of 
VV and AA since they are not independent from the other polarizations considered above.
Note that Eq.~\eqref{eq:chisqcontact} goes to zero when the theoretical cross section vanishes (SM limit) as one would expect.

In the light $M_X$ scenario (see Sec.~\ref{subsection:lightMX})
in which one of the new neutral gauge boson is too light and invalidates 
the effective contact interaction approach,
it is mandatory to recast the LEP-II constraints for the 
contact interactions into the cross section level to do the analysis. 
We checked that for the heavy $M_X$ scenario, using either the effective contact interaction or cross section approach
give the same results.

\subsection{Drell-Yan Constraints at the LHC}

In this section we recap the experiments of the Drell-Yan cross section 
for SM $Z$-boson and heavy $Z^\prime$ at the LHC.

\subsubsection{$Z$-boson on-shell decay at the LHC}
\label{subsection:LHCZData}
By using the measurement of the Drell-Yan cross section for the $Z$-boson production, 
the properties of the $Z$ are well determined at the LHC. 
Among all the final states of the $Z$-boson decay, the dilepton signature 
is the most relevant to distinguish signal from background. 
It is commonly believed that the Drell-Yan constraint 
$q\bar{q}\to Z \to l^+l^-$ from the LHC is weaker than LEP EWPT data 
because of the relative larger uncertainties from the hadronic background
than the QED background.
However, to be careful, we first check a direct Drell-Yan constraints from the LHC~\cite{Aaboud:2017buh}. 
The data of electron-positron pair ($ee$) and muon-pair ($\mu\mu$) final states 
are given by Tables 3 and 4 respectively in Ref.~\cite{Aaboud:2017buh}. 
In the signal region located around $Z$-boson mass (the invariant mass $80<m_{ll}/\gev<120$), 
we found that the systematic uncertainties of Drell-Yan background is larger 
than the data statistic uncertainties in both $ee$ or $\mu\mu$ final state.
We have also checked that the EWPT constraints in Table~\ref{table:ewobs} are much stronger than 
LHC Drell-Yan constraint.
%

\begin{figure}[h!]
    \subfloat[
    ]{
        \includegraphics[width=0.3\textwidth]{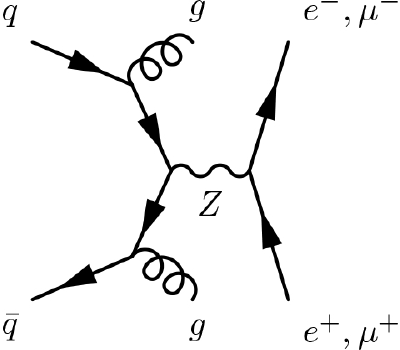}
        \label{fig:QCDzjj}
    }\hspace{5mm}
    \subfloat[
    ]{
        \includegraphics[width=0.3\textwidth]{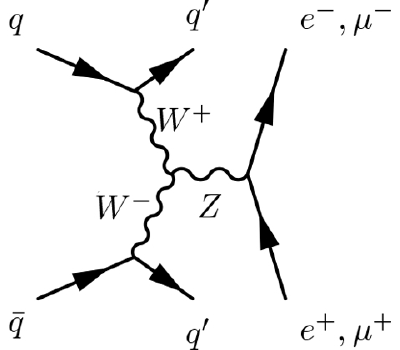}
        \label{fig:VBFzjj}      
    }
 \caption{The Feynman diagrams of electroweak Drell-Yan process. 
  }
 \label{fig:zjj}
 \end{figure}

On the other hand, $Z$-boson can be singly produced either by 
radiation from the incoming partons (Fig.~\ref{fig:QCDzjj}) or 
$t$-channel exchange of a $W$ gauge boson (Fig.~\ref{fig:VBFzjj}). 
To constrain the G2HDM modified $Zl^+l^-$ couplings, the later process is more 
useful than the former because QCD processes
usually suffer from larger systematical uncertainties than the electroweak ones.
Recently, ATLAS~\cite{Aaboud:2017emo} reported a 
fiducial electroweak cross section of $\sigma^{Zjj}_{\rm{EW}}=119\pm 16 \pm 20\pm 2$ fb and 
$\sigma^{Zjj}_{\rm{EW}}=34.2\pm 5.8 \pm 5.5\pm 0.7$ fb for 
dijet invariant masses $m_{jj}$ greater than $250\gev$ and $1\tev$, respectively. 
The SM simulated cross sections $\sigma^{Zjj}_{\rm{EW}}({\rm SM})$ are also given in Table 5 of Ref.~\cite{Aaboud:2017emo}, 
where central values and the uncertainties are given as $125.2\pm 3.4$ fb for $m_{jj}>250\gev$ and $38.5\pm 1.5$ fb for $m_{jj}> 1\tev$.   

Comparing with the SM, except for the $Z{l^+l^-}$ couplings, the G2HDM did not 
modify much of the cross section. Namely, the electroweak cross section of the G2HDM version can be 
simply rescaled as 
\begin{equation}
\sigma^{Zjj}_{\rm{EW}}({\rm G2HDM})=\sigma^{Zjj}_{\rm{EW}}({\rm SM})\times 
\mathcal{R} \, ,
\end{equation}
where
\beq
\mathcal{R} =
\left [
\frac{\mathcal{C}^{ZWW}_{\rm{G2HDM}}}
{\mathcal{C}^{ZWW}_{\rm{SM}}}
\right ]^2 \frac{{\rm BR}^{\rm{G2HDM}}_{Z \to f\bar{f}}}{{\rm BR}^{\rm{SM}}_{Z \to f\bar{f}}}
= {\cal O}_{11}^2 \frac{{\rm BR}^{\rm{G2HDM}}_{Z \to f\bar{f}}}{{\rm BR}^{\rm{SM}}_{Z \to f\bar{f}}} \, ,
\label{Rfract}
\eeq
and $f=e,\mu$. However, similar to direct Drell-Yan $Z$ boson search, we found that the value of 
$\mathcal{R}$ is not easy to derivate from unity and 
the power of constraining the parameter space in G2HDM is not as strong as LEP EWPT constraints.

Finally, we have numerically verified that the allowed G2HDM parameter space 
is hardly changed at all whether the direct and electroweak Drell-Yan $Z$ boson constraints at the LHC 
are included or not. Again, this is because both constraints at the LHC are much weaker than LEP EWPT constraints. 
Hence, we will not take into account the LHC Drell-Yan constraints 
from the on-shell $Z$ decay in our numerical works so as to save some computer resources.

\subsubsection{LHC $Z'$ Boson Search at High-mass Dilepton Resonances}

The Drell-Yan constraints can also be very powerful for the new gauge bosons in G2HDM 
once they can be singly produced~\cite{Huang:2017bto}. 
Unlike the study in Ref.~\cite{Huang:2017bto} 
which only $W^{'3}$-like $Z_i$ is considered, 
we extend it here to any $Z_i$ with all the possible composition. 
Recently, ATLAS collaboration~\cite{Aad:2019fac} reported a new 
result on  dilepton resonances with the integrated luminosity of
$139$~fb$^{-1}$ and the center-of-mass energy $\sqrt{s}=13\tev$. 
They indicated that the lower limit on the mass of $Z'$ boson for a simplified model can be raised up to $4-5\tev$.
Considering this new measurement, we update the constraints of the heavy neutral gauge boson 
masses in G2HDM and the upper limits of $g_H$ and $g_X$.

In Fig. 3 of Ref.~\cite{Aad:2019fac}, one can see the upper limits of 
cross section times branching ratio BR($Z' \to l^+ l^-$) are based on the 
ratio of the total width $\Gamma_{Z'}$ of $Z'$ divided by its $M_{Z'}$. 
Depending on this ratio, the limits can be altered by a factor of $\sim 5$.  
As shown in Appendix~\ref{appendix:decaywidths}, 
the $\Gamma_{Z_i}/M_{Z_i}$ in the G2HDM shall be always less than $0.06$.
Hence, for a conservative approach, we can simply apply the ATLAS result 
by using their upper limit associated with $\Gamma_{Z^\prime}/M_{Z^\prime}=0.06$.

Furthermore, the $Z_i$ total decay width relies on whether $Z_i$ decays to the new particles in G2HDM. 
The heavy new fermions in G2HDM are assumed to be very heavy so that 
they do not affect the EW-scale physics in any significant way. 
On the other hand, the $Z_i$ invisible decay to a scalar DM pair can be a more important channel because 
the upper limits of various parameters 
can be weaker than the one without taking into account the $Z_i$ decays to the DM pair. 
The openings of the scalar channels  as well as other  
channels with one vector and one scalar particles in the final states of $Z_i$ decay
makes the parameter spaces of  the gauge and scalar sectors entangle with each other.
Thus a complete analysis becomes quite formidable.
In Eq.~\eqref{eq:ZptoDD}, one can see that the invisible decay width  of $Z_i$
has two different limits, $M_D\ll M_{Z_i}$ for maximum invisible decay and 
$M_D>M_{Z_i}$ for zero invisible decay.  
For the sake of simplicity, we will be contented by presenting the results based on these two benchmark invisible decay widths. 
In this study, we adopt $M_D=M_{Z_i}/10$ for maximum invisible decay but 
we found that the $\Gamma(Z_i\to D D^*)$ can differ within an accepted range of $\sim 6\%$ 
comparing with the massless $M_D$ case.

Using \texttt{MadGraph5}~\cite{Alwall:2014hca}, we compute 
the cross section $\sigma(p p\to Z_i)$. 
Since we enforce that the cross section is computed at the resonance, 
we only used a minimum cut given by the default parameter card in \texttt{MadGraph5}. 
It is very \texttt{CPU} time consuming to estimate the cross section 
point by point throughout all the parameter space.   
Nevertheless, the cross section can be obtained by simply rescaling the vector and axial vector couplings
$v^i_f$ and $a^i_f$ using Eqs.~\eqref{eq:vecfacmy0} 
and~\eqref{eq:axifacmy0}. Hence, by using the same reasoning as before we include the latest ATLAS $Z'$ limit 
in our scan by using the following chi-squared function
\begin{equation}
\chi^2_{\rm{ATLAS}} = 2.71 \times
\left[\frac{\sigma_{\text{G2HDM}}(pp\to Z_i)\times {\rm BR}_{\text{G2HDM}}(Z_i\to l^+l^-) }
{\sigma^{\rm{95\%}}_{\text{ATLAS}}(pp\to Z')\times {\rm BR}^{\rm{95\%}}_{\text{ATLAS}}(Z'\to l^+l^-) }\right]^2 \; , 
\label{eq:chisqatlas}
\end{equation} 
where the branching ratio ${\rm BR}_{\text{G2HDM}}(Z_i\to l^+l^-)$ can be found in Appendix~\ref{appendix:decaywidths} 
and $\sigma^{\rm{95\%}}_{\text{ATLAS}}(pp\to Z')\times {\rm BR}^{\rm{95\%}}_{\text{ATLAS}}(Z'\to l^+l^-)$ 
is $95\%$ C.L. taken from the curve associated with $\Gamma_{Z'}/M_{Z'}=0.06$ 
in Fig.~3 of Ref.~\cite{Aad:2019fac}.   

\section{Results} \label{section:results}

\subsection{Numerical Methodology\label{subsection:methodology}}

Our aim is to determine the $68\%$ and $95\%$ allowed parameter space of the G2HDM 
which are favored by all of the experimental data presented in the previous 
section. In this paper, we will use the profile-likelihood (PL) method to perform the statistical data analysis. 
We recap the PL method in the following. 
Briefly, the PL method is a well popular statistical method to deal with the 
multi-dimensional parameter space which treats the unwanted parameters as nuisance parameters.
In other words, if a proposed model has $n$-dimensional parameter space and 
we are only interested in $p$ of those dimensions, 
then the PL method can remove the unwanted $n-p$ dimensions which we are not interested in, 
by maximizing the likelihood over them.   

There are 4 new parameters in the gauge sector of G2HDM. They are
the two new gauge couplings $g_H$ and $g_X$ and the two new scales
$v_\Phi$ and $M_X$. 
Our results will be presented in two-dimensional parameter regions with 68\% and 95\% confidence levels (C.L.).
Take the plane ($g_H,g_X$) as an example. 
After marginalizing over the other two parameters $v_\Phi$ and $M_X$,
an integral of the likelihood function $\mathcal{L}(g_H,g_X)$ can be written as 
\begin{equation}
\frac{\int_{\mathcal{C}}\mathcal{L}(g_H,g_X)dg_H dg_X}
{\rm{normalization}}
=\varrho \; , 
\end{equation}
where $\mathcal{C}$ is the smallest area bound with a fraction $\varrho$ of the total probability 
and the normalization in the denominator is the total probability with $\mathcal{C}\to\infty$. 

The total $\chi^2_{\rm Total}(g_H,g_X,v_\Phi,M_X)$ we will use in our analysis is the sum of Eqs.~(\ref{eq:chisqewobs}), (\ref{eq:chisqcontact}), 
and (\ref{eq:chisqatlas}), namely
\begin{equation}
\chi^2_{\rm Total} = \chi^2_{Z-{\rm pole}} + \chi^2_{\rm LEP-II} + \chi^2_{\rm ATLAS}  \; ,
\end{equation}
where we have suppressed the arguments of all the $\chi^2$ functions.
We adopt the statistical sensitivity as
\begin{equation}
\Delta \chi^2 = \chi^2_{{\rm Total}}- {\rm min} \, (\chi^2_{{\rm Total}})\; .
\end{equation}
Since our likelihood is modeled as a pure Gaussian distribution, 
{\it i.e.} $\mathcal L\propto \exp (- \chi^2/2)$,
one can connect the $\chi^2$ to the confidence level:  
the 68\%~(95\%) C.L. in a two dimensional parameter space corresponding to 
$\Delta\chi^2=-2\ln(\mathcal{L}/\mathcal{L}_{\rm{max}})=2.30~(5.99)$. 
Here $\mathcal{L}_{\rm{max}}$ is the maximum value of the likelihood in the region $\mathcal{C}$. 

There are two interesting scenarios: (i) heavy $M_X$ and (ii) light $M_X$. 
The heavy $M_X$ scenario will result in two new heavy neutral gauge bosons 
$Z_2 \equiv Z^\prime$ and $Z_3 \equiv Z^{\prime\prime}$, 
and the measured boson located at $Z$-pole will be the lightest one, $Z_1 \equiv Z$. 
However, the light $M_X$ scenario will result in a new boson $Z_1$ 
lighter than the $Z$-pole which is usually called dark $Z$ ($Z_D$) 
or dark photon ($\gamma_D$). In this case, $Z_2$ corresponds to the $Z$-pole $Z_2 \equiv Z$
and $Z_3 \equiv Z^\prime$.
Hence, we choose our $M_X$ scan ranges for two scenarios, 
\begin{equation}
\frac{M_X}{\tev} : \left\{
\begin{array}{l}
 \displaystyle  \left[ 0.1:10\right] \quad ({\rm heavy}~M_X) \\[3mm]
 \displaystyle  \left[ 10^{-6}:0.08\right] \quad ({\rm light}~M_X)
  \end{array}
  \right. \; .
\label{eq:mxprior}  
\end{equation}
For the other three parameters, we use the same ranges for the two scenarios of $M_X$~\footnote{
There is also the possibility of both $M_X$ and $v_\Phi$ are light, which may lead to $Z_3 = Z$ 
and both $Z_1$ and $Z_2$ are lighter than $Z$. We will reserve this interesting scenario in future work.},  
\begin{eqnarray}
10^{-8} \leq & g_H &\leq g^{\texttt{SM}} = \frac{e^{\texttt{SM}}}{\sin\theta_W} = 0.65 \; , \nonumber \\
10^{-8} \leq & g_X &\leq g^{\prime \texttt{SM}} = \frac{e^{\texttt{SM}}}{\cos\theta_W} = 0.35 \; ,  \\
5\tev \leq & v_\Phi &  \leq 200\tev  \; .\nonumber
\end{eqnarray}  

We perform random scans by using MultiNest v2.17~\cite{Feroz:2008xx} 
with 30000 living points, an enlargement factor reduction parameter 0.5 and a stop tolerance factor $10^{-3}$ . 
For sampling coverage, we combined several scans and finally obtained 
$\sim 10^{5}$ samples for each scenario.

\subsection{Heavy $M_X$ Scenario\label{subsection:heavyMX}}

\begin{figure}
    \subfloat[
    ]{
        \includegraphics[width=0.5\textwidth]{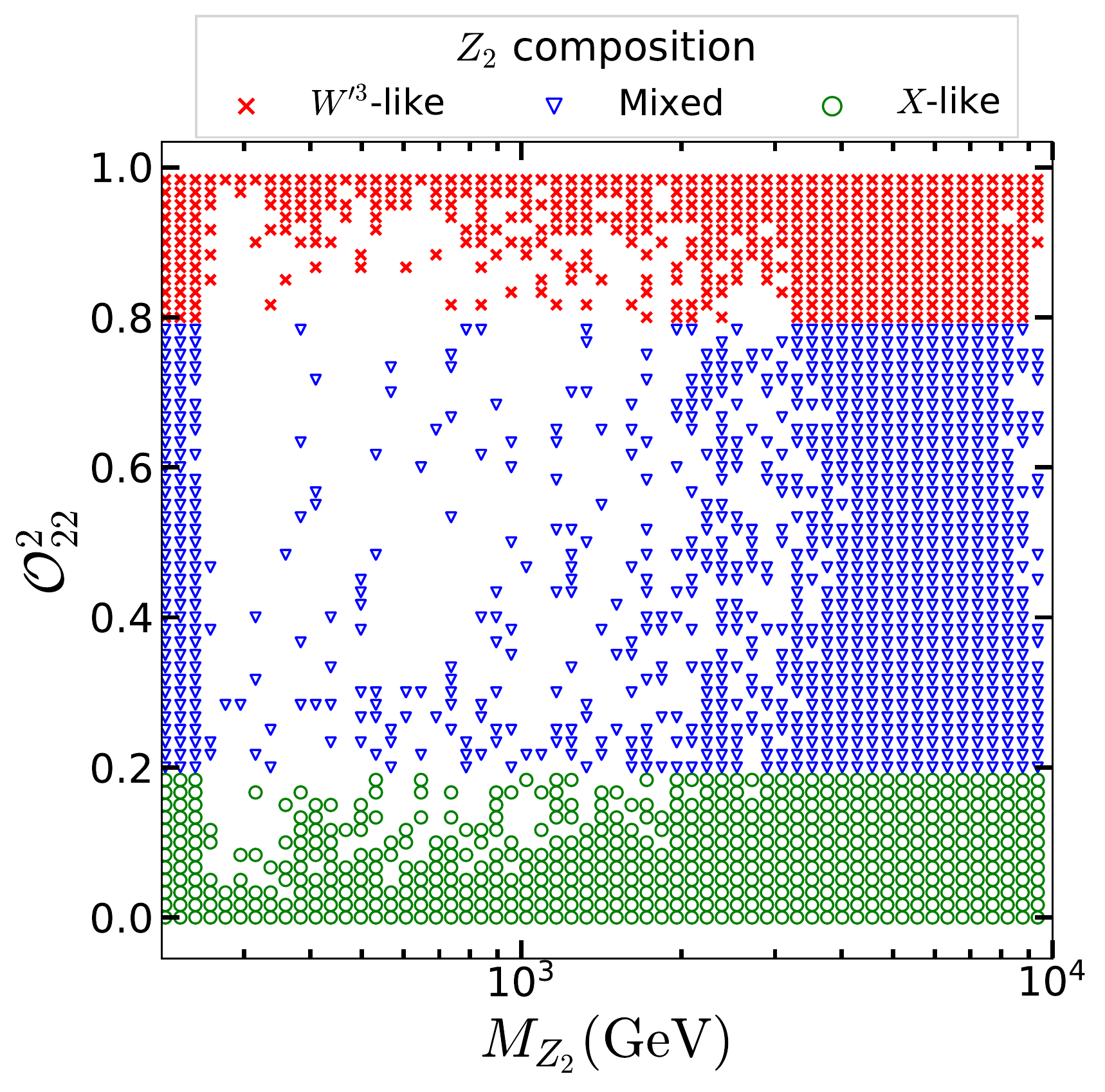}
        \label{heavyMX_mz2_o22}
    }
    \subfloat[
    ]{
        \includegraphics[width=0.5\textwidth]{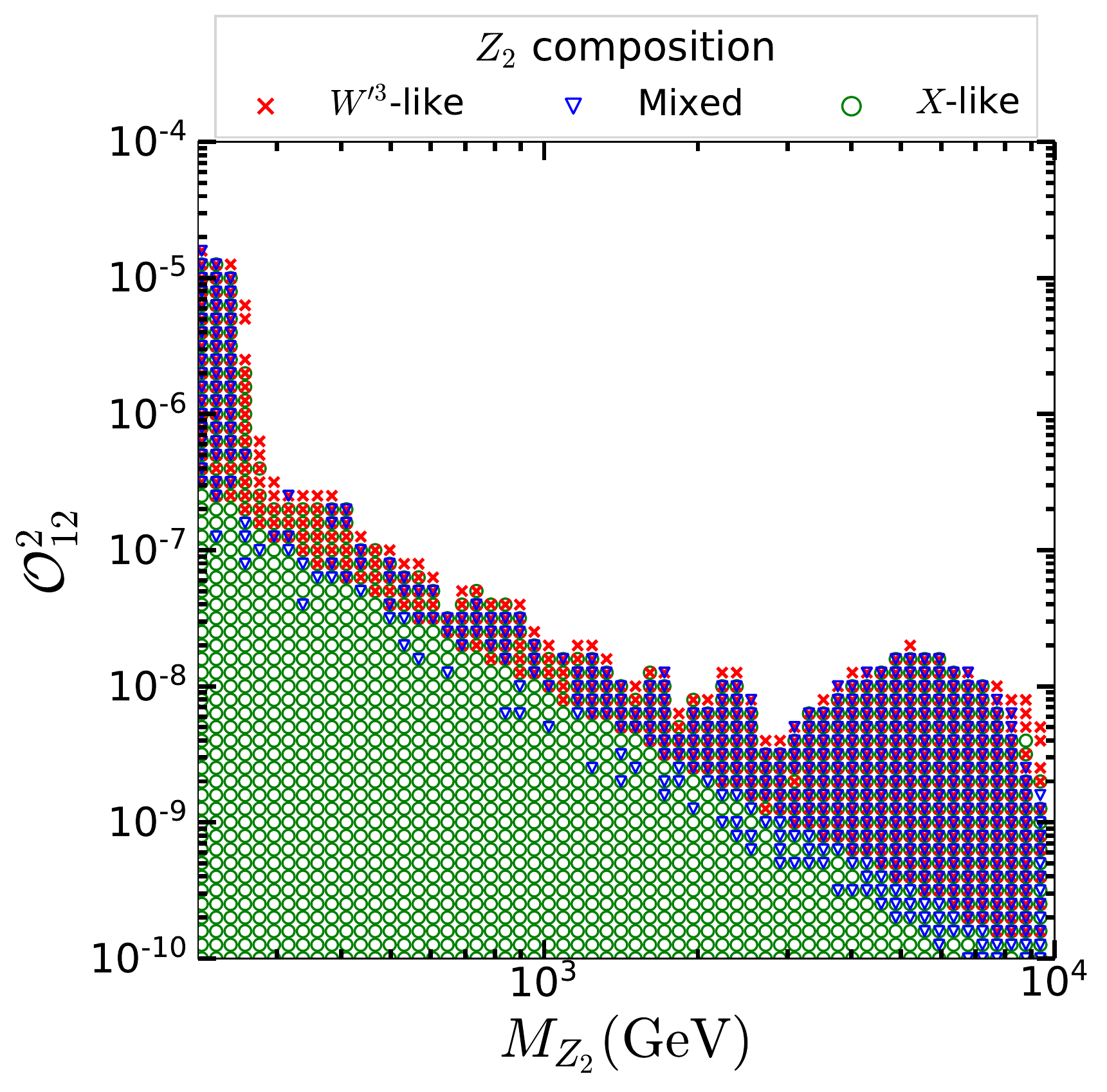}
        \label{heavyMX_mz2_o12}
    }
 \caption{ \small Scatter plots in 1$\sigma$ on (a) ($M_{Z_2}$, ${\cal O}^2_{22}$) plane 
 and (b)  ($M_{Z_2}$, ${\cal O}^2_{12}$) plane for the heavy $M_X$ scenario.
 The red cross region with ${\cal O}_{22}^2$ between 0.8 and 1.0 represents the points of
 $W^{\prime3}$-like $Z_2$ boson; the blue triangle region with ${\cal O}_{22}^2$ between 0.2 and
 0.8 represents the points mixed with $W^{\prime 3}$, and the green circle
 region with ${\cal O}_{22}^2$ between 0.0 and 0.2 represents the points of $X$-like $Z_2$ boson.
 }
 \label{heavyMX_mzp_oi2}
 \end{figure}
 
In the heavy $M_X$ scenario, the mass of $Z_1$ boson is located at around $Z$-pole ($\sim 91\gev$)
so that $Z_1$ is identified as the SM $Z$-boson. 
Note that $Z_1(Z)$ boson physics is strongly affected by the
different composition of $Z_2$ ($Z^\prime$) but not the heaviest boson $Z_3$ ($Z^{\prime\prime}$) 
because $Z_3$ is heavier than $Z_2$ in our parameter choices and therefore has less impact.

In Fig.~\ref{heavyMX_mzp_oi2}, we present the scatter points of the composition of 
$Z_2 =  \mathcal O_{12} Z^{\rm SM} + \mathcal O_{22} W'^3 +  \mathcal O_{32} X$
for the $1\sigma$ region based on the likelihoods 
described in Sec.~\ref{subsection:methodology}. 
The color code hereafter represents the three different composition of $Z_2$. 
Recalling Eq.~(\ref{eq:Zcomposition}), we define 
$W'^{3}$-like $Z_2$ with condition $\mathcal{O}_{22}^2 > 0.8$ (red crosses {\color{red}$\times$}),
mixed state $Z_2$ with $0.2<\mathcal{O}_{22}^2 <0.8$ (blue triangles {\color{blue}$\triangledown$}),
and $X$-like $Z_2$ with condition $\mathcal{O}_{22}^2 <0.2$ (green circles {\color{green}$\circ$}).

The $1\sigma$ allowed scatter points projected on the 
($M_{Z_2}$,~$O^2_{22}$) and ($M_{Z_2}$,~$O^2_{12}$) planes
are depicted in Figs.~\ref{heavyMX_mz2_o22} and \ref{heavyMX_mz2_o12},
respectively.  
From the density of distribution in Fig.~\ref{heavyMX_mz2_o22}, 
we can clearly see that the mixed state $Z_2$ (blue triangles) is less evenly distributed 
because it needs some trade-off between the two new gauge couplings $g_H$ and $g_X$. 
In Fig.~\ref{heavyMX_mz2_o12}, we projected the same parameter space on 
the plane ($M_{Z_2}$, $\mathcal{O}_{12}^2$). 
Note that the mixing $\mathcal{O}_{12}^2$ presents how $Z_2$ is consisted of $Z^{\rm{SM}}$. 
Therefore, very small $\mathcal{O}_{12}^2$ implies $\mathcal O_{32}^{2} \approx (1- \mathcal O_{22}^2)$
from the orthogonality of $\mathcal O$.
Furthermore, the upper limit of $v_\Phi$ sets an lower
limit of the $\mathcal{O}_{12}^2$ for the red cross region.
If $v_\Phi$ goes to infinity, $Z_2$ becomes super heavy and decouple. 
The composition of $Z^{\rm{SM}}$ in $Z_2$ should then be negligible, thus $\mathcal{O}_{12}^2$ vanishes
in this limit. 
We note that the excluded concave up region of $M_{Z_2}$ between 250 GeV and 6 TeV 
on the upper limit of ${\cal O}_{12}^2$
is due to the constraint from ATLAS $Z'$ search.

\begin{figure}
    \subfloat[
    ]{
        \includegraphics[width=0.5\textwidth]{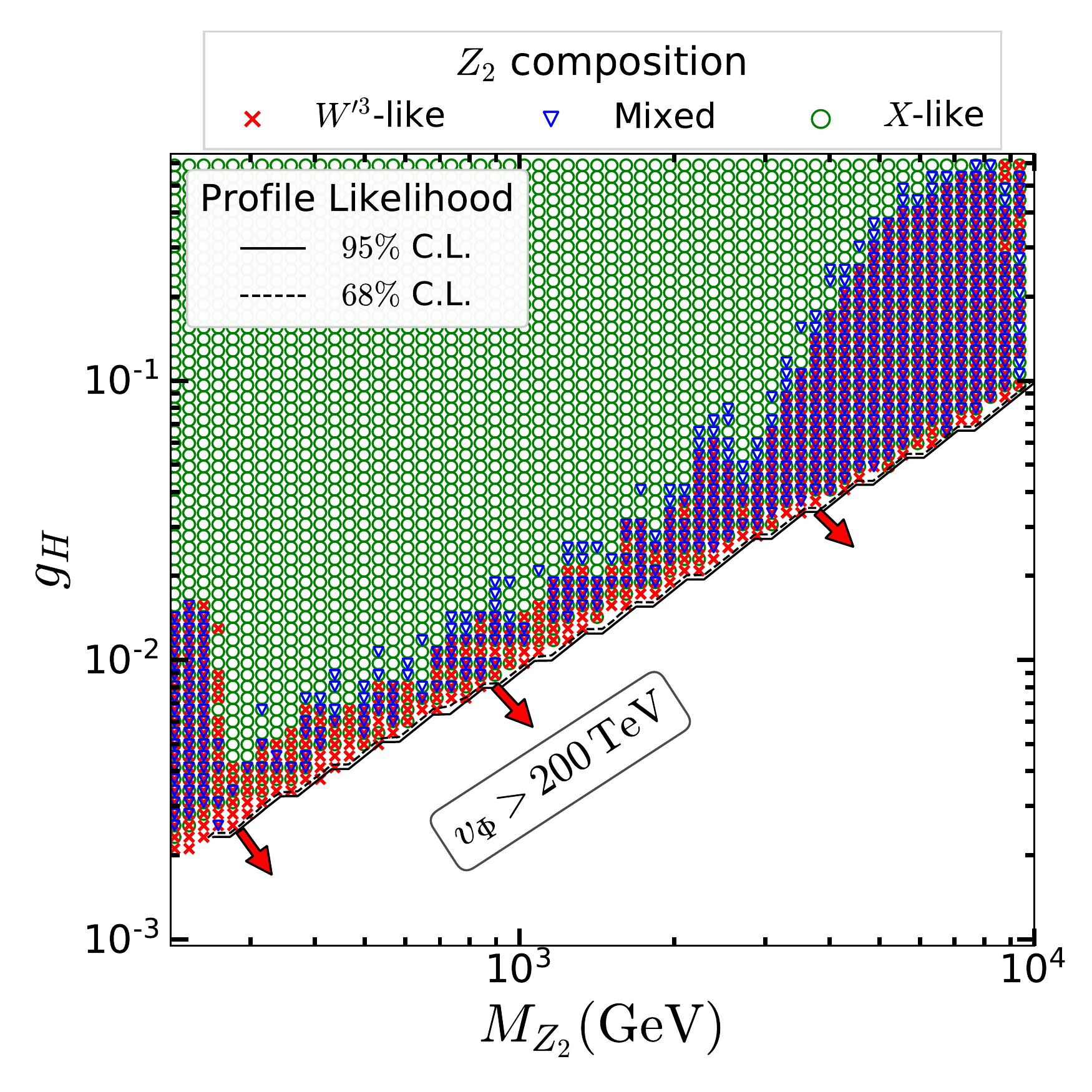}
        \label{heavyMX_mz2_gH_like}
    }
    \subfloat[
    ]{
        \includegraphics[width=0.5\textwidth]{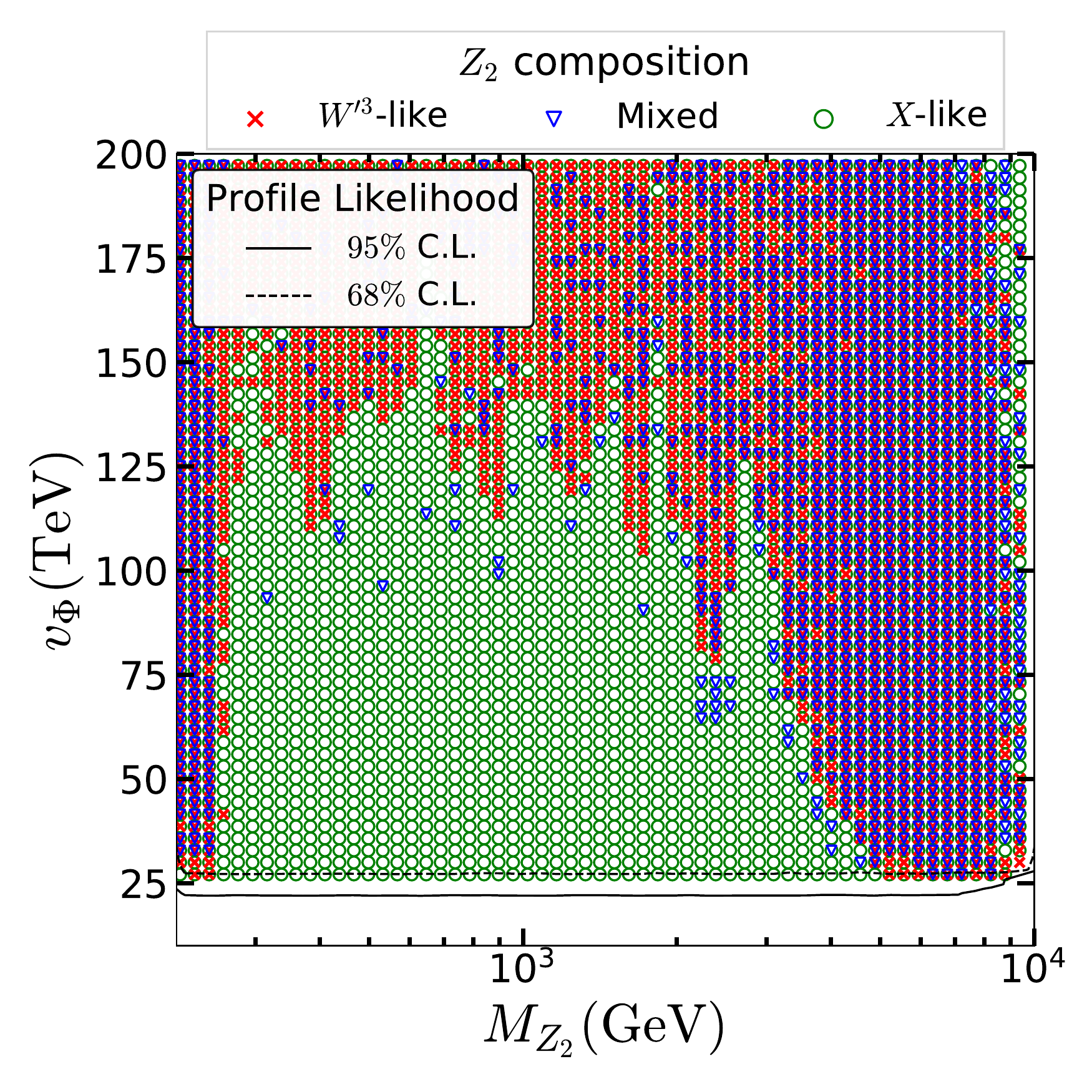}
        \label{heavyMX_mz2_vp_like}      
    }
 \caption{Scatter plots in $1\sigma$ on (a) ($M_{Z_2}$, $g_{H}$) plane and (b) ($M_{Z_2}$, $v_{\Phi}$) plane
 for the heavy $M_X$ scenario.
The color code is the same as Fig.~\ref{heavyMX_mzp_oi2}. 
The $1\sigma$ and $ 2\sigma$ contours of the profile likelihood are also shown.
}
 \label{heavyMX_mz2vpgH_like}
 \end{figure}
 

In Fig.~\ref{heavyMX_mz2vpgH_like}, we show the $1\sigma$ (dashed) and 
$2\sigma$ (solid) likelihood contours with scatter points inside the $1\sigma$ region on the 
(a) ($M_{Z_2}$, $g_{H}$) and (b) ($M_{Z_2}$, $v_\Phi$) planes.
In Fig.~\ref{heavyMX_mz2_gH_like}, we can see that the $W^{\prime 3}$-like red
crosses form a band with a tendency proportional to $g_H$. 
This is because for a $W^{\prime 3}$-like $Z_2$, 
$m^2_{Z_2} \approx g_H^2 (v^2 + v^2_\Phi)/4 \approx
g_H^2 v^2_\Phi/4$ which can be extracted from the (3,3) element of the mass 
matrix in Eq.~\eqref{eq:MgaugeSMrot}.
We can also see that at the lower bound of this band, the 95\% and 68\% C.L. contours
are overlapped because this lower bound is 
due to our choice of $v_\Phi<200\tev$ in its upper scan range, 
not from the likelihood results.
This implies that in the upper edge of this red band where $g_H$ has larger value, 
the value of $v_\Phi$ there is smaller.
Therefore, the upper bound of this red cross band corresponds to the lower values of $v_\Phi$, 
which can be excluded by the $\chi^2$ tolerance as we can see 
in Fig.~\ref{heavyMX_mz2_vp_like} where the scatter plot is projected on the ($M_{Z_2}$, $v_\Phi$) plane.
Surprisingly, in Fig.~\ref{heavyMX_mz2_gH_like}, the blue triangle band,
corresponding to mixing mostly between $W^{\prime 3}$ and $X$ bosons, matches the red cross
band. This can be understood as the mass of $Z_2$ being dominated by the
(3,3) element of Eq.~\eqref{eq:MgaugeSMrot} even for an $80\%$ $X$ boson composition. 
In the same figure, we can see the green circles running from below the two
red cross and blue triangle bands up to the upper limit of $g_H$. In
other words, we can see how the $M_{Z_2}$ passes from being dominated by the
(3,3) element of Eq.~\eqref{eq:MgaugeSMrot} (red crosses), which is $g_H$-dependent,
to being dominated by the $g_H$-independent (4,4) element (green circles).

One particular feature of Fig.~\ref{heavyMX_mz2_vp_like} is that the low
$M_{Z_2}$ and low $v_\Phi$ region (lower left corner) is covered only with $X$-like points while
both $W^{\prime 3}$-like and mixed points only approach this corner down to a
curved bound. 
This curved section in the lower bound can be related to the
curved upper bound for $W^{\prime 3}$ and mixed points in
Fig.~\ref{heavyMX_mz2_gH_like} for low $M_{Z_2} < 200$ GeV and $g_H\lesssim 10^{-2}$.
These curves in the upper bound (Fig.~\ref{heavyMX_mz2_gH_like})
and in the lower bound (Fig.~\ref{heavyMX_mz2_vp_like}) can be understood as smaller $v_\Phi$
requiring larger $g_H$ to pass EWPT. In particular, if $g_H$ is small,
$v_\Phi$ has to be large in order to have a sizable diagonal (3,3) element in
 the mass matrix in Eq.~\eqref{eq:MgaugeSMrot}, while 
the off-diagonal (2,3) and (3,2) elements remain small.
However, the mixing effects from the off-diagonal elements 
are not negligible and expected to be stronger when the ${Z_2}$ mass is getting closer to the ${Z^\text{SM}}$ mass. 
This gives rise to the upper and lower bounds that we see in Figs.~\ref{heavyMX_mz2_gH_like} and
\ref{heavyMX_mz2_vp_like}, respectively, for the $W^{\prime 3}$-like points.
Such behaviour is not displayed for the $X$-like points since they do not depend strongly on $g_H$.

The ATLAS $Z'$ constraint almost rules out the region $250\gev<M_{Z_2}< 6\tev$
for $W'^{3}$-like and mixed $Z_2$, except the region with $v_\Phi>100\tev$.
However, the $X$-like $Z_2$ at the same region has not been affected much by the ATLAS $Z'$ constraint.

\begin{figure}
    \subfloat[
    ]{
        \includegraphics[width=0.5\textwidth]{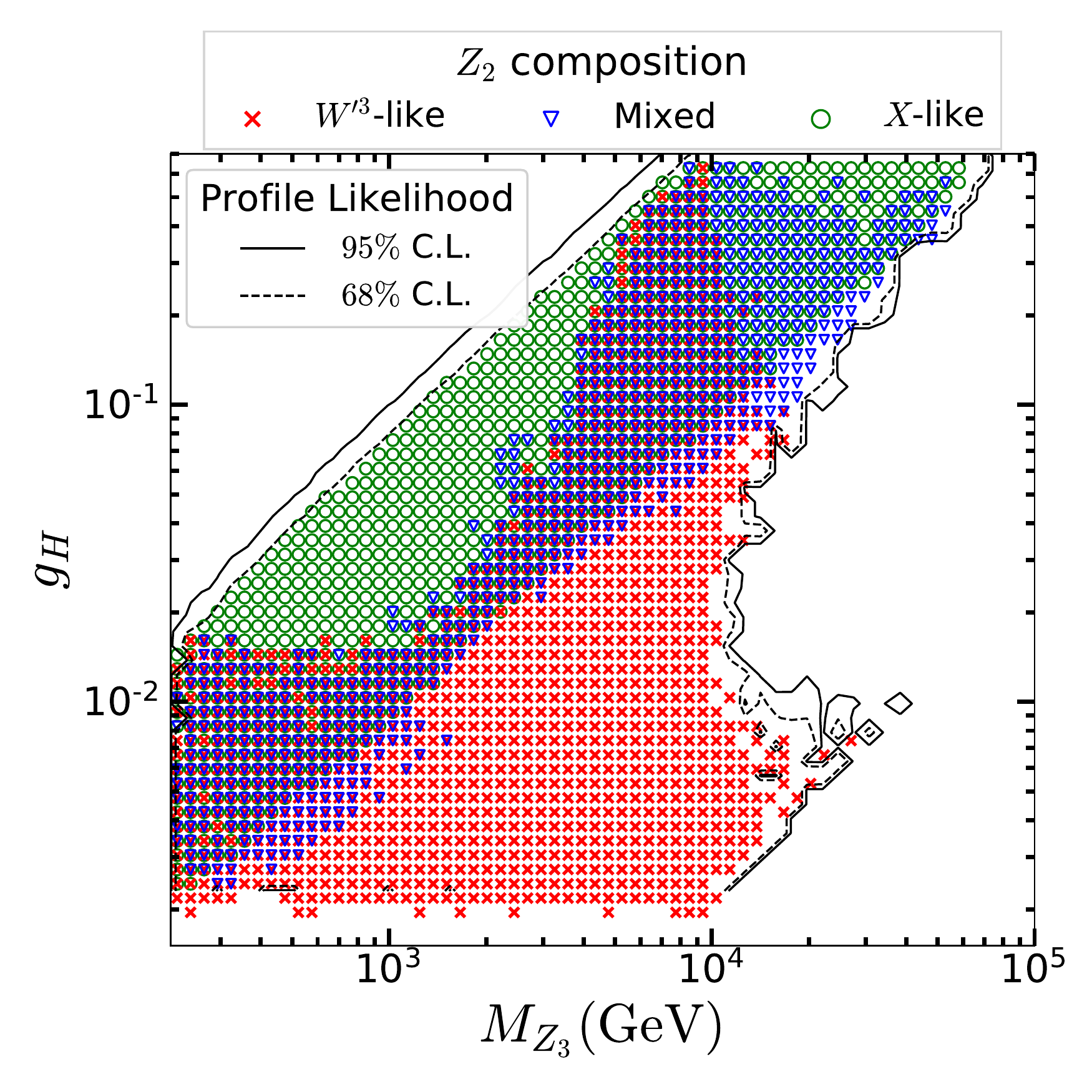}
        \label{heavyMX_mz3_gH_like}
    }
    \subfloat[
    ]{
        \includegraphics[width=0.5\textwidth]{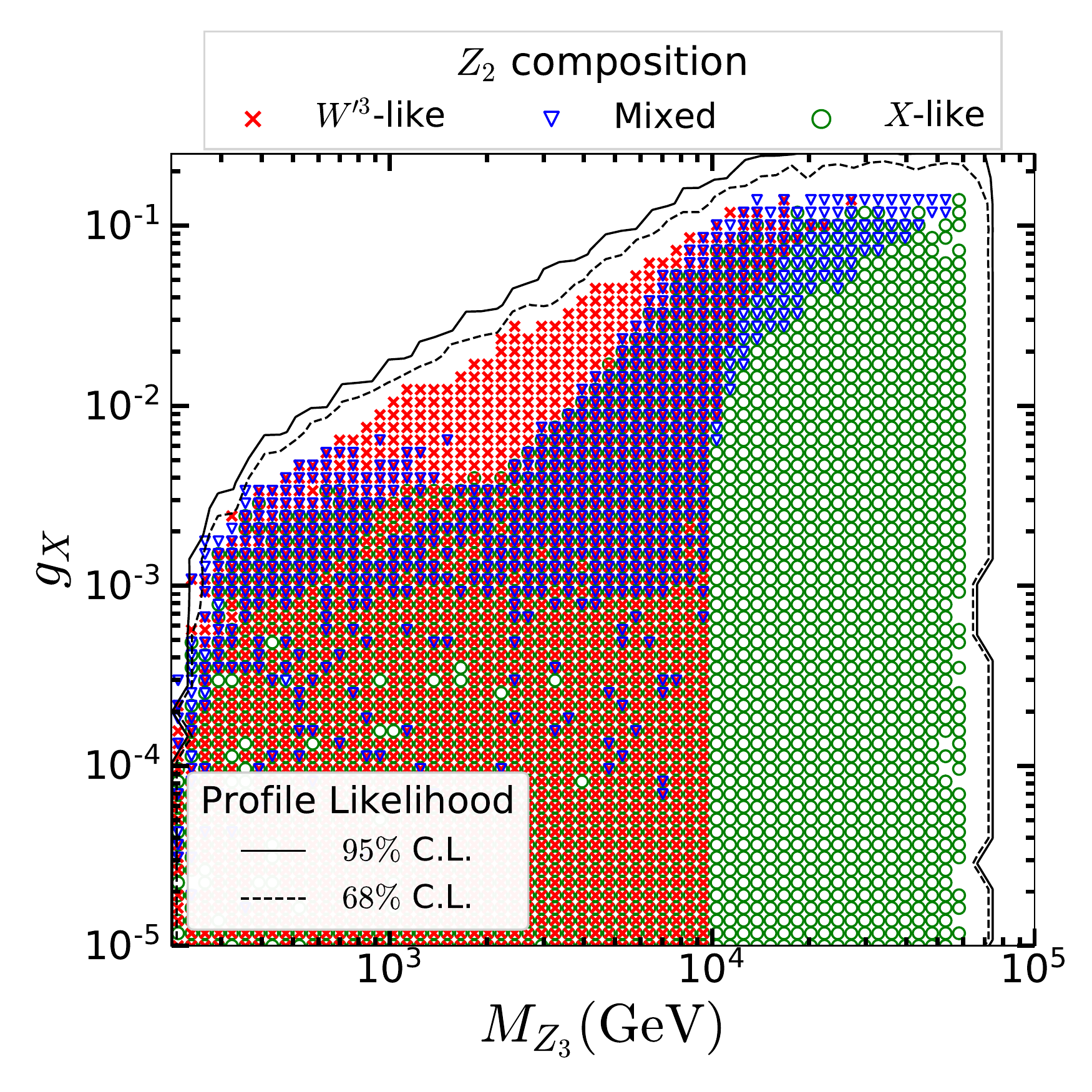}
        \label{heavyMX_mz3_gX_like}      
    }
 \caption{Scatter plots in $1\sigma$ on (a) ($M_{Z_3}$, $g_{H}$) plane and (b) ($M_{Z_3}$, $g_{X}$) plane
 for the heavy $M_X$ scenario.
 The color code is the same as Fig.~\ref{heavyMX_mzp_oi2}.
 The $1\sigma$ and $ 2\sigma$ contours of the profile likelihood are also shown.
  }
 \label{heavyMX_mz3gHgX_like}
 \end{figure}

Similarly, in Fig.~\ref{heavyMX_mz3gHgX_like}, we show the $1\sigma$ (dashed)
and $2\sigma$ (solid) likelihood contours with scatter points inside the
$1\sigma$ region on the (a) ($M_{Z_3}$, $g_{H}$) and (b) ($M_{Z_3}$, $g_{X}$)
planes. From Fig.~\ref{heavyMX_mz3_gH_like}, one can easily see that the
$X$-like $Z_2$ boson (green circles) forms a band whose tendency is
proportional to the $g_H$. This can be understood by the fact that the
composition of the $Z_3$ in this case is mainly from $W'^{3}$, which has a mass
proportional to  $0.5 \,g_H \sqrt{v^2 +v_{\Phi}^2} \approx 0.5\,g_H v_{\Phi} $ again coming
from the (3,3) element of the mass matrix in Eq.~\eqref{eq:MgaugeSMrot}.  
On the other hand, in the case of the
$W'^{3}$-like $Z_2$ boson (red crosses), the mass of the $Z_3$ almost does not
depend on $g_H$. Indeed, the composition of $Z_3$ is now mainly from $X$ and
$M^2_{Z_3} \approx (g_X^2 (v^2 + v_\Phi^2) + M_X^2)$. This is clearly shown in
Fig.~\ref{heavyMX_mz3_gX_like}, when $g_X$ is small ($g_X <
3\times10^{-3}$),
the mass of $Z_3$ in the red cross region is dominated by
$M_X$ and less than our set-up limit of $10^{4}$~GeV.
However, when $g_X$ is getting bigger, the mass of the $Z_3$ can
be dominated by the $g_X v_\Phi$ term for sufficiently large value of
$v_\Phi$.  We can also see that the EWPT data sets upper bounds on $g_H$ and
$g_X$.  
The excluded concave up region of $250\gev<M_{Z_2}< 6\tev$ in
	Fig.~\ref{heavyMX_mz3_gH_like} for the $W'^3$-like and mixed composition
	of $Z_2$ is again due to the ATLAS $Z'$ search which does not apply for
	the $X$-like case.  As a result, the ATLAS $Z'$ search cannot constrain on
$g_X$ for $W'^3$-like points as clearly shown in
Fig.~\ref{heavyMX_mz3_gX_like}. 

\subsection{Light $M_X$ Scenario\label{subsection:lightMX}}

\begin{figure}
    \subfloat[
    ]{
        \includegraphics[width=0.5\textwidth]{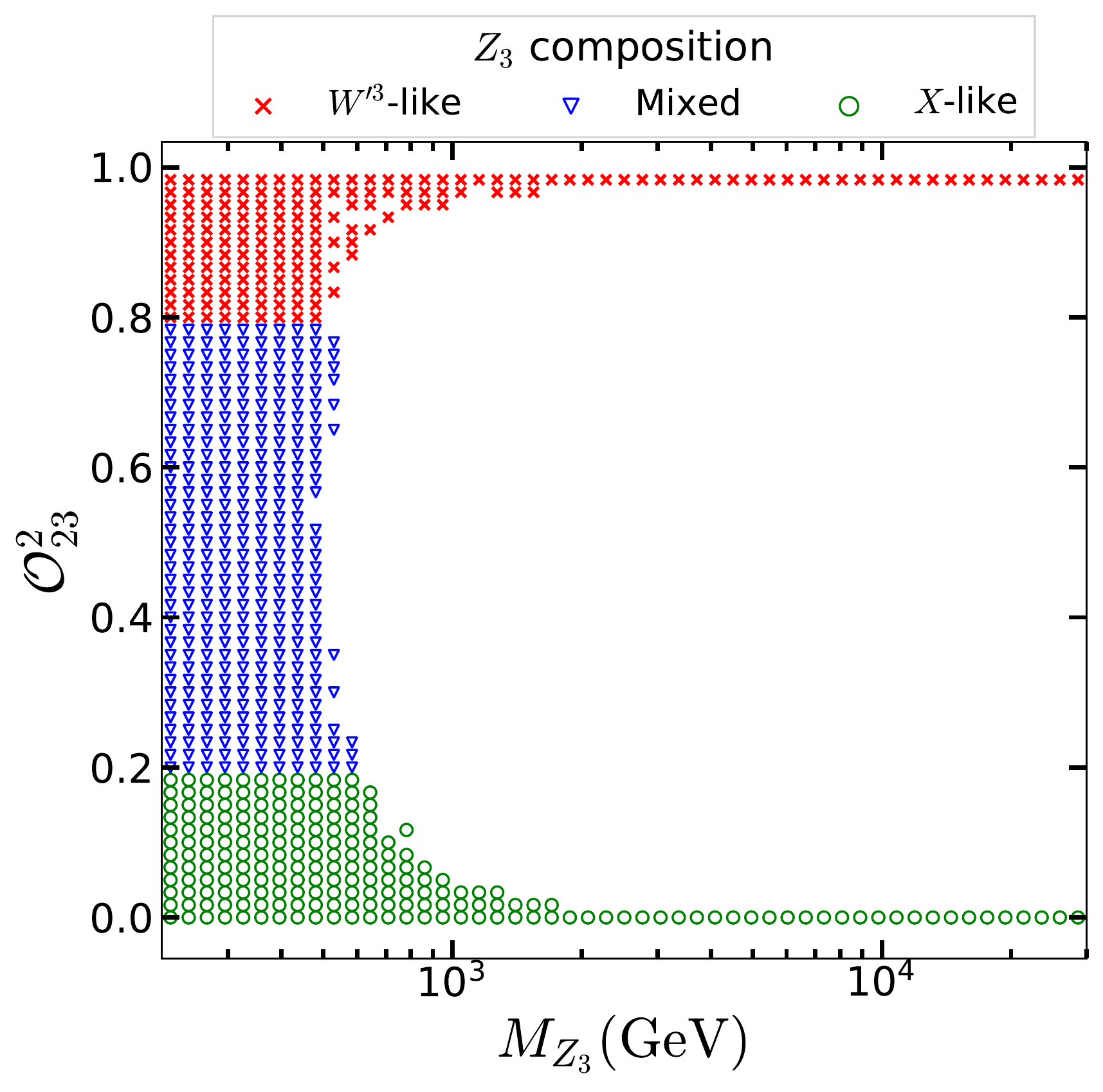}
        \label{lightMX_mz3_o23}      
    }
    \subfloat[
    ]{
        \includegraphics[width=0.5\textwidth]{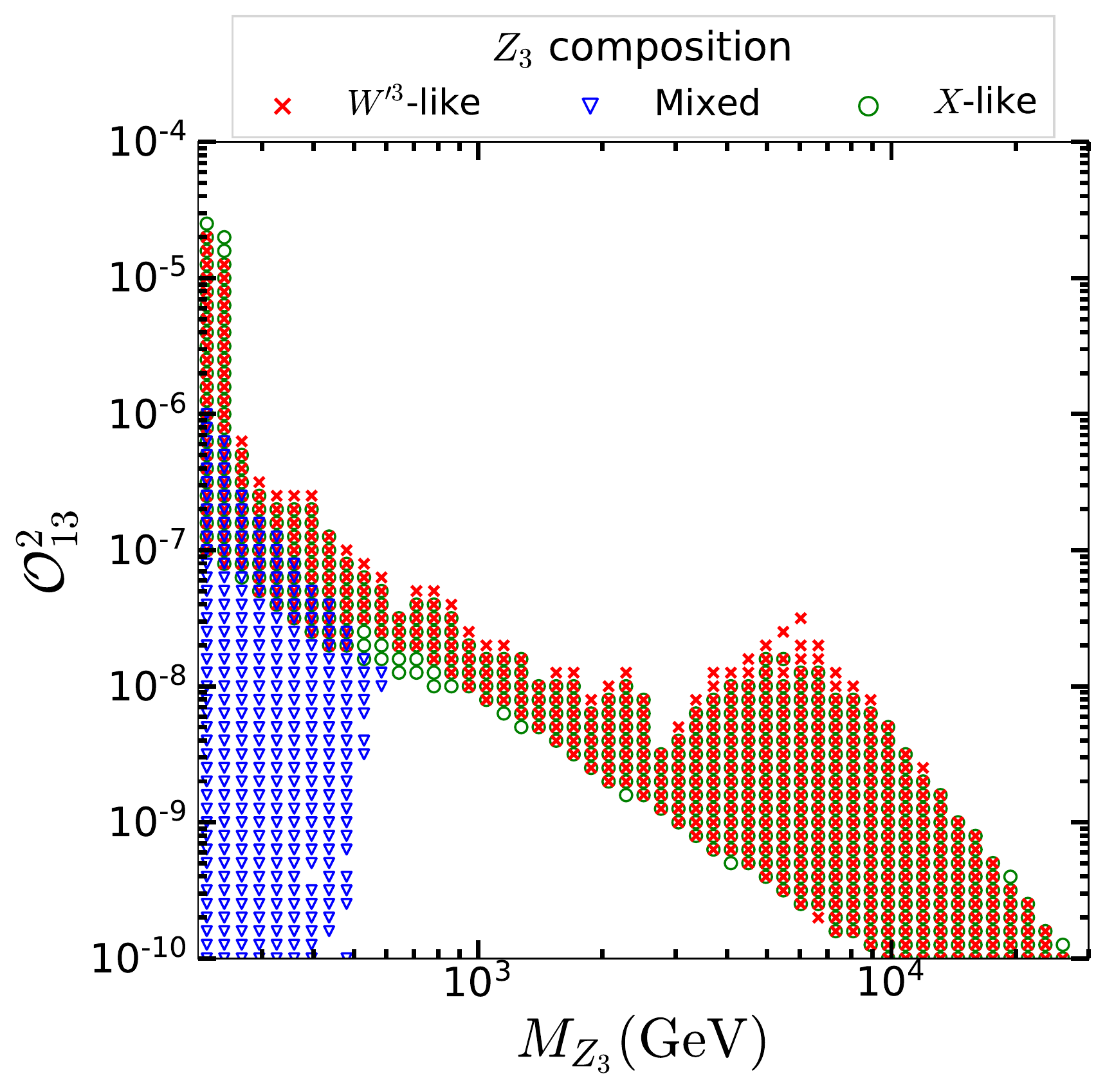}
        \label{lightMX_mz3_o13}
    }
\caption{
Scatter plots in $1\sigma$ region on 
(a) ($M_{Z_3}$, $O^2_{23}$) plane and 
(b) ($M_{Z_3}$, $O^2_{13}$) plane
for the light $M_X$ scenario.
The red cross represents the points of $W^{\prime3}$-like $Z_3$ boson, the
blue triangle represents the points mixed states ($Z^{\rm SM}$, $W^{\prime3}$ and
$X$) $Z_3$ boson, and the green circles represents $X$-like $Z_3$ boson. 
}
\label{lightMX_mzp_oi3}
\end{figure}

In the light $M_X$ scenario, we require that the mass of $Z_2$ boson is always at 
around $Z$-pole ($\sim 91\gev$).  
In this scenario, the lightest $Z_1$ with mass less than the $Z$-boson mass can be 
the dark photon or dark $Z$, while the conventional $Z'$ is the heaviest boson $Z_3$. 
We note that the composition of $Z_3$ is given by 
$Z_3 = \mathcal O_{13} Z^{\rm SM} + \mathcal O_{23} W'^3 +  \mathcal O_{33} X$.
The $1\sigma$ allowed scatter points projected on the 
($M_{Z_3}$,~$O^2_{23}$) and ($M_{Z_3}$,~$O^2_{13}$) planes
are depicted in Figs.~\ref{lightMX_mz3_o23} and \ref{lightMX_mz3_o13},
respectively.  
The color code for the composition of $Z_3$ is the same as 
in Fig.~\ref{heavyMX_mzp_oi2} for $Z_2$.

\begin{figure}
    \subfloat[
    ]{
        \includegraphics[width=0.5\textwidth]{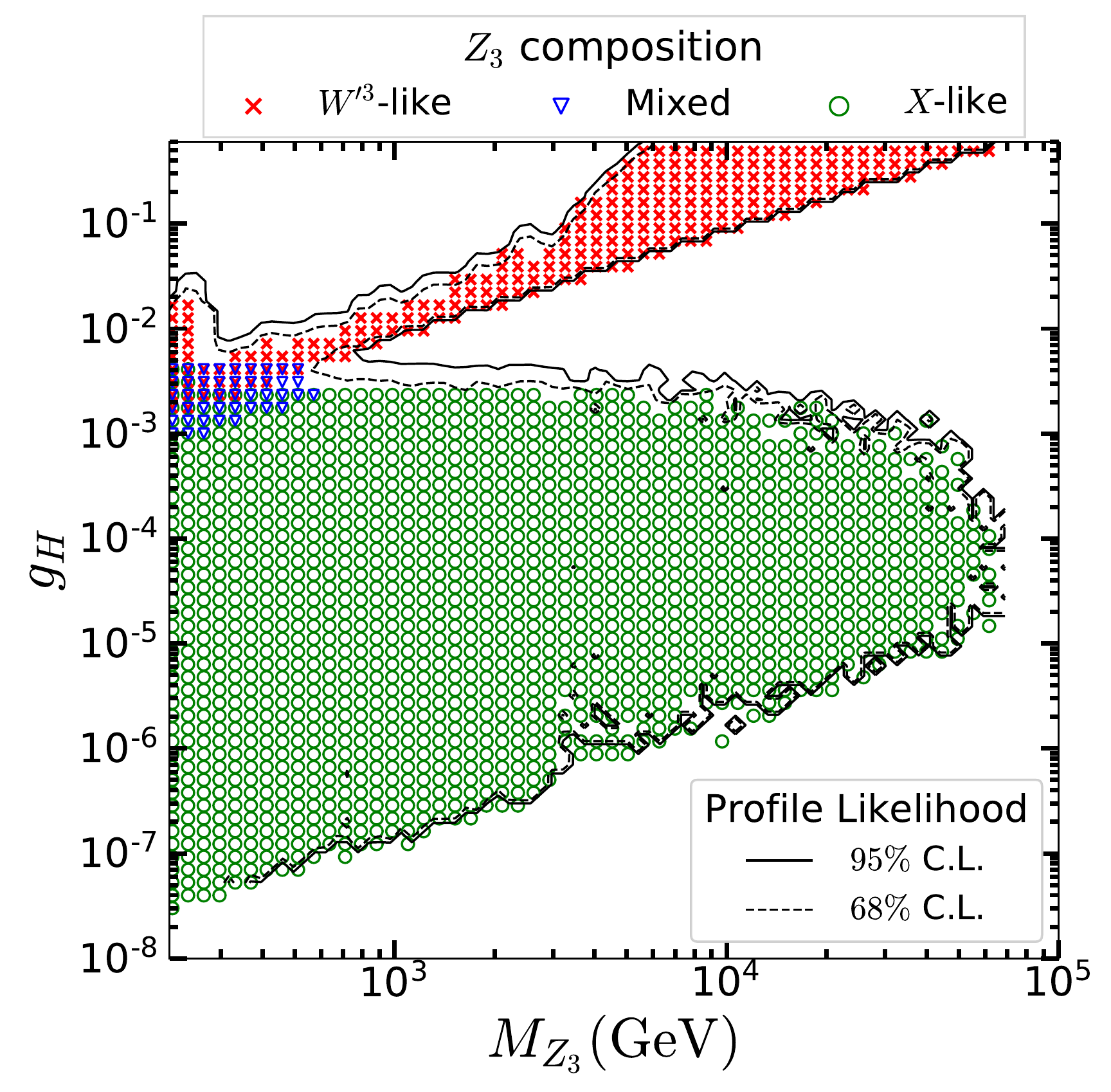}
        \label{lightMX_mz3_gH_like}
    }
    \subfloat[
    ]{
        \includegraphics[width=0.5\textwidth]{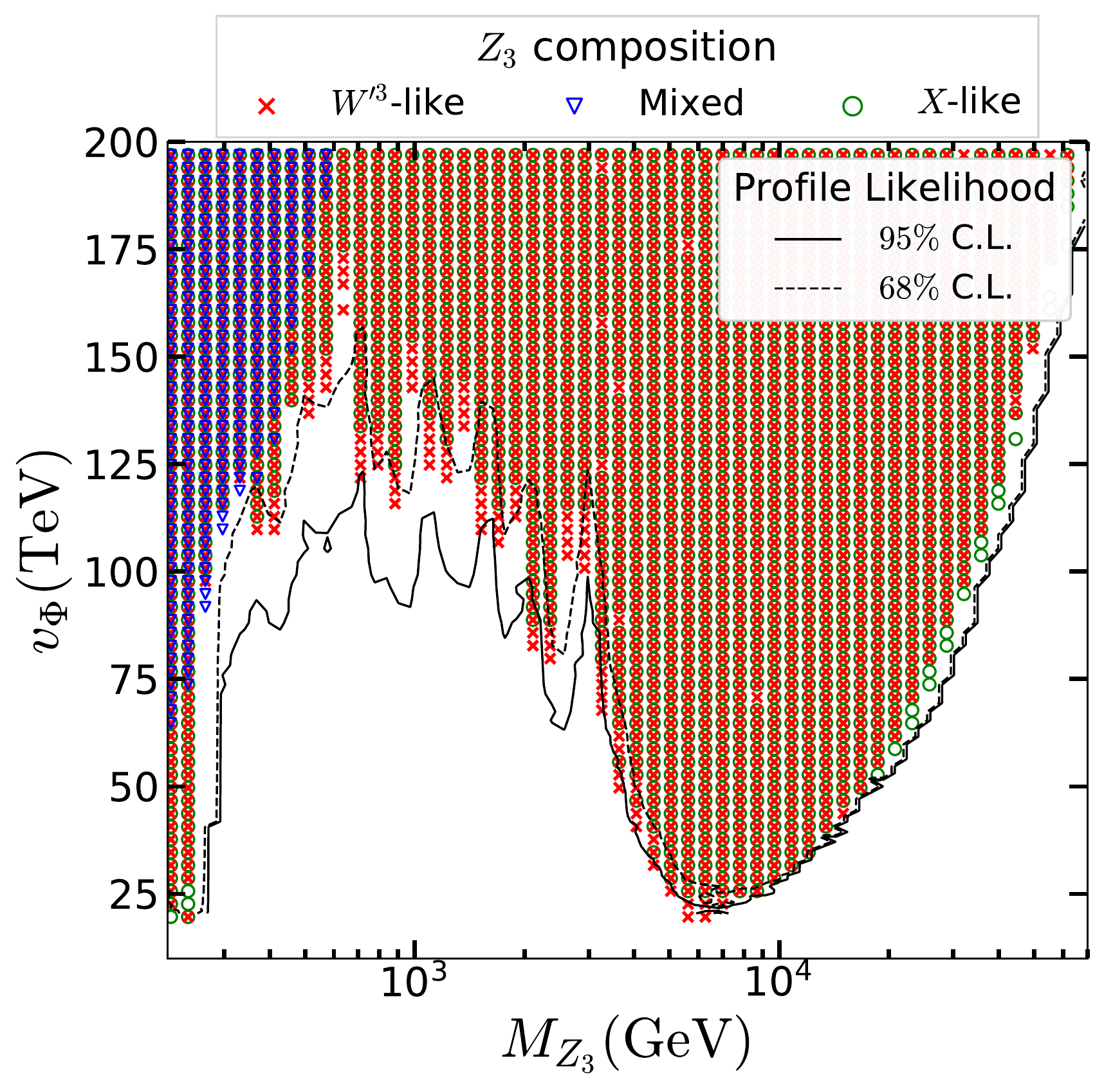}
        \label{lightMX_mz3_vp_like}      
    }
 \caption{\small Scatter plots in $1\sigma$  on (a) ($g_{H}$ $M_{Z_3}$) plane and (b) ($v_\Phi$, $M_{Z_3}$) plane 
 for the light $M_X$ scenario.
 The markers are the same as Fig.~\ref{lightMX_mzp_oi3}.
 The $1\sigma$ and $ 2\sigma$ contours of the profile likelihood are also shown.
 }
 \label{lightMX_mz3vpgH_like}
 \end{figure}

An obvious feature of Fig.~\ref{lightMX_mz3_o23} is that 
the mixed state of $Z_3$ (blue triangles) has a mass upper limit. 
Intuitively, it requires some trade-off between the gauge couplings $g_H$ and $g_X$ 
which results in $M_{Z_3}\lesssim 500\gev$. 
This effect will be discussed with more detail later in Fig.~\ref{lightMX_mz3vpgH_like}. 
In Fig.~\ref{lightMX_mz3_o13}, we can see that the $Z^{\rm{SM}}$ composition of $Z_3$ is again small. 
However, unlike the heavy $M_X$ scenario, 
the $X$-like $Z_3$ boson has a similar distribution as 
$W'^3$-like $Z_3$ boson. 
Additionally, the mixed $Z_3$ state at the mass region between $210\gev$ and $700\gev$ 
cannot be excluded by the ATLAS $Z'$ constraint which is also different from the heavy $M_X$ scenario.

In analogous to Fig.~\ref{heavyMX_mz2vpgH_like}, we show in Fig.~\ref{lightMX_mz3vpgH_like}
the $1\sigma$ (dashed) and $2\sigma$ (solid) likelihood 
contours with scatter points in the $1\sigma$ region on the (a) ($M_{Z_3}$, $g_{H}$) and (b) ($M_{Z_3}$, $v_\Phi$) planes.
Comparing Figs.~\ref{heavyMX_mz2_gH_like} and \ref{lightMX_mz3_gH_like},
we have a clear separation between the $W^{\prime 3}$-like (red crosses) and $X$-like (green circles) 
regions in this light $M_X$ scenario. 
As before, the $W^{\prime 3}$-like red crosses follow a tendency proportional to
$g_H$ again because of the dominance of the (3,3) element of
Eq.~\eqref{eq:MgaugeSMrot} in $M_{Z_3}$, {\it i.e.}, $M_{Z_3}\approx g_Hv_\Phi/2$. 
Other features shared between $W^{\prime 3}$-like points in
Figs.~\ref{heavyMX_mz2_gH_like} and \ref{lightMX_mz3_gH_like} are the
distribution of $v_\Phi$ values; 
the $g_H$ lower bound owes to $v_\Phi$ upper bound
but its upper bound owes to $\chi^2$ tolerance. 
As expected, the ATLAS $Z'$ search can constrain $g_H$ and $v_\Phi$ at the mass region $250\gev<M_{Z_2}< 6\tev$. 
However, the gauge coupling for $X$-like $Z_3$ is proportional to $g_X$ 
not $g_H$ so that the ATLAS $Z'$ search cannot constrain on $g_H$ at the $X$-like region, 
indicated by green circles.
The $X$-like region in Fig.~\ref{lightMX_mz3_gH_like} has a $g_H$ upper bound
around $10^{-2}$ given by the $\chi^2$ tolerance and most likely related to
the lower bound on $v_\Phi$ displayed on Fig.~\ref{lightMX_mz3_vp_like}. The
mass of $Z_3$, $M_{Z_3}$, in this $X$-like green region can be approximated by $\sqrt{g^2_X
v^2_\Phi + M_X^2}$, this is why there is not a clear $g_H$ dependence as in
the $W^{\prime 3}$-like points. 
In Fig.~\ref{lightMX_mz3_gH_like}, as one would expect, the mixed region
corresponds approximately to the intersection between $X$-like and $W^{\prime
3}$-like regions, extending lightly into their exclusive regions.  This means
that the upper and lower bound of the mixed region are approximately given by
the upper bound of the $X$-like region and the lower bound of the $W^{\prime
3}$-region, respectively. If we increase our maximum $v_\Phi$ value, the lower
bound of the $W^{\prime 3}$-like region would reach lower $g_H$ and the
maximum $M_{Z_3}$ for the mixed region would be increased. This is more clear
after looking at Fig.~\ref{lightMX_mz3_vp_like} where the maximum $M_{Z_3}$
value for the three regions grows with the value of $v_\Phi$.

\begin{figure}
    \subfloat[
    ]{
        \includegraphics[width=0.5\textwidth]{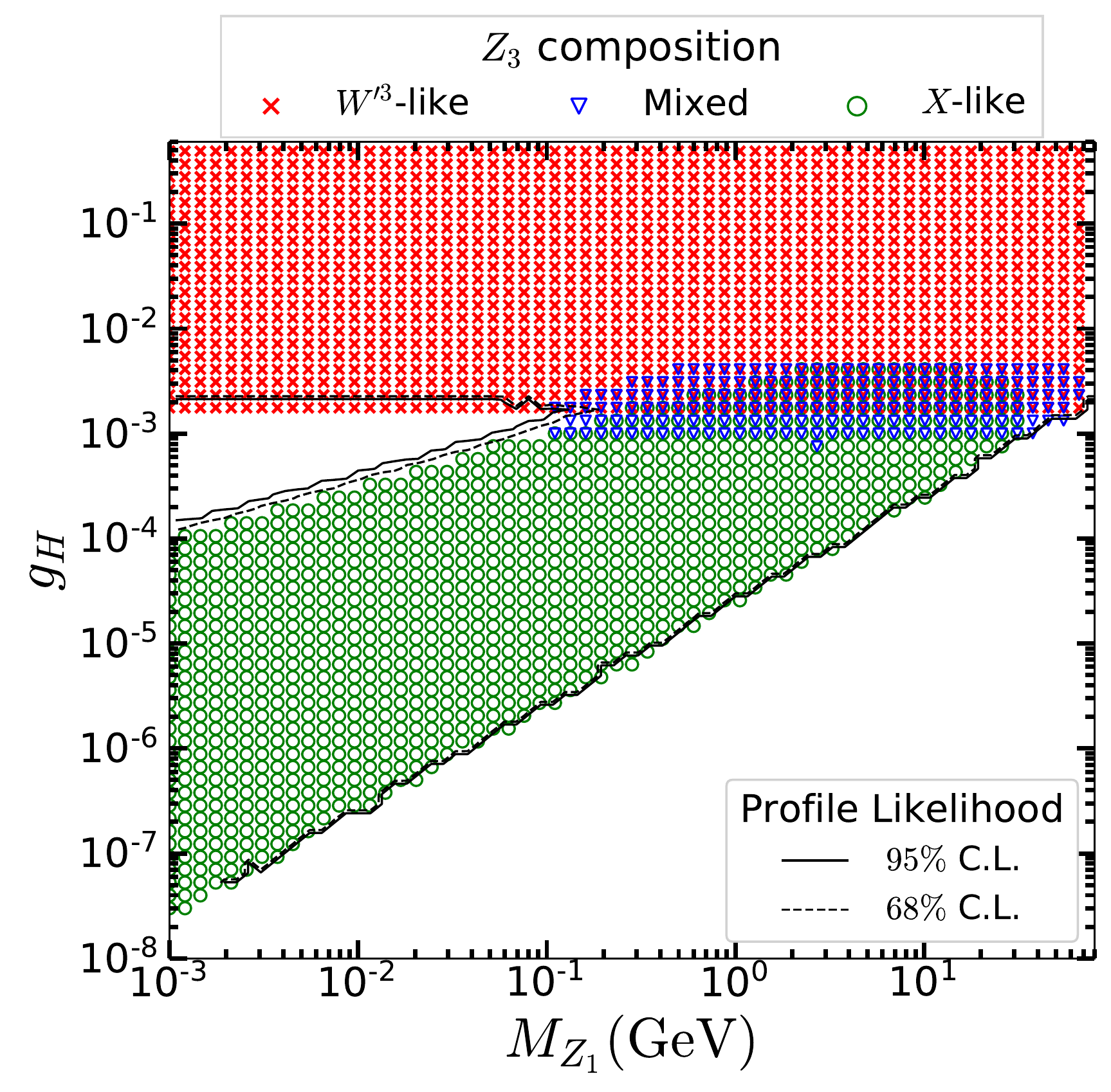}
        \label{lightMX_mz1_gH_like}
    }
    \subfloat[
    ]{
        \includegraphics[width=0.5\textwidth]{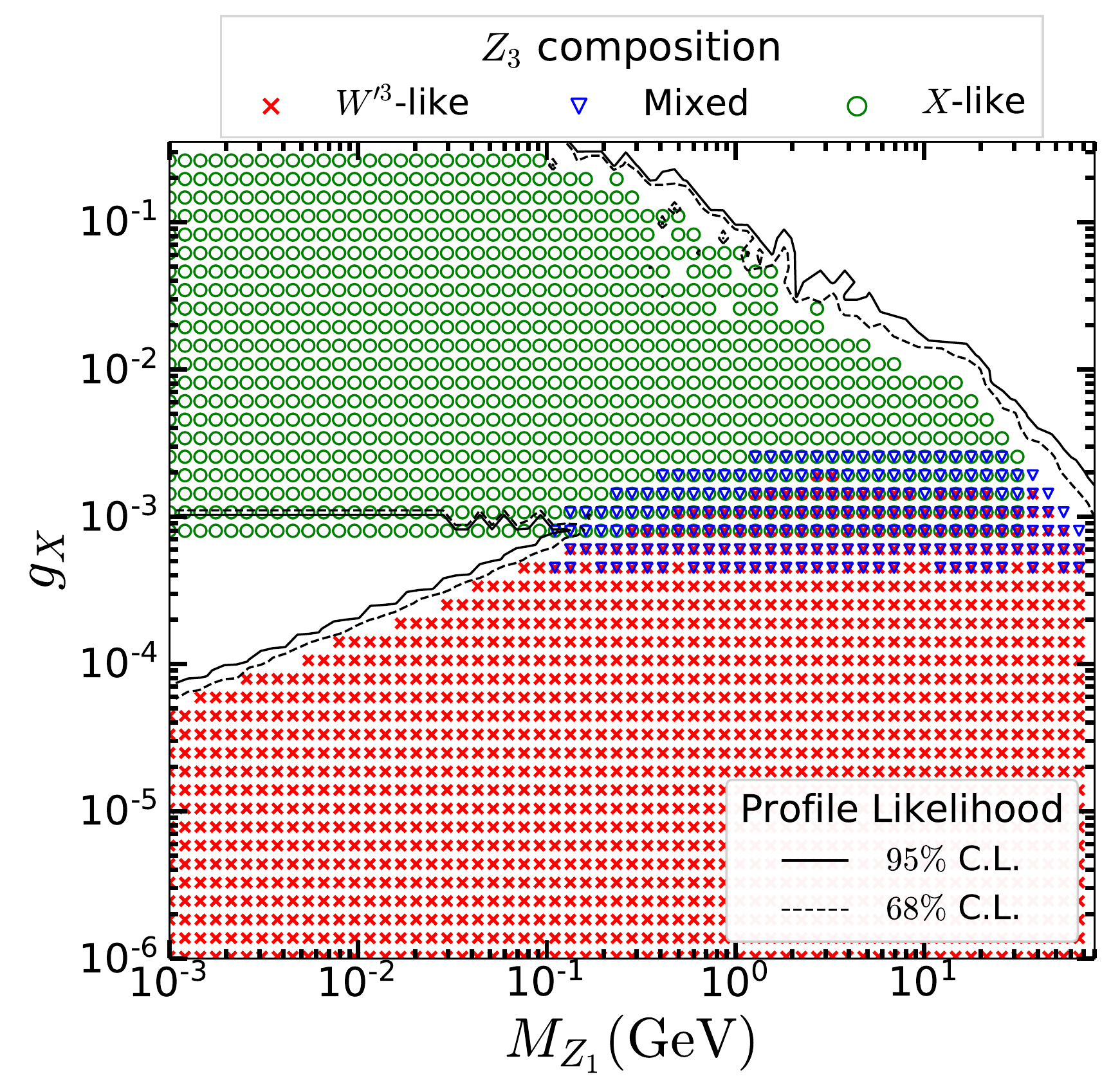}
        \label{lightMX_mz1_gX_like}      
    }
 \caption{\small Scatter plots in $1\sigma$ on (a) ($g_{H}$ $M_{Z_1}$) plane and (b) ($g_X$, $M_{Z_1}$) plane for the light
 $M_X$ scenario.
 The markers are the same as Fig.~\ref{lightMX_mzp_oi3}.
 The $1\sigma$ and $ 2\sigma$ contours of the profile likelihood are also shown.
 }
 \label{lightMX_mz1gHgX_like}
 \end{figure}

Similarly, in Fig.~\ref{lightMX_mz1gHgX_like}, we show the $1\sigma$ (dashed) and $2\sigma$ (solid) likelihood contours with scatter points in the 
$1\sigma$ region on the (a) ($M_{Z_1}$, $g_H$) plane and (b) ($M_{Z_1}$, $g_X$) planes. Again, the red cross represents the points of $W^{\prime3}$-like $Z_3$ boson, 
the blue triangle represents the points of mixed state ($Z^{\rm SM}$, $W^{\prime3}$ and $X$) $Z_3$ boson, and the green circle represents $X$-like $Z_3$ boson. We note that in this scenario, $Z_1$ is considered as a dark photon~\footnote{The lightest boson $Z_1$ can be tested in the dark photon experiments 
but it is beyond the scope of this work. We will return to this in the future.} 
and has mass range from 1 MeV to $Z$-pole. One can easily see that the $Z_3$ composition 
is clearly separated on the planes of ($M_{Z_1}, g_H$) and ($M_{Z_1},g_X$). 
In particular, 
while the $W'^3$-like $Z_3$ boson parameter space is distributed in the region of larger $g_H$ and smaller $g_X$,
the $X$-like $Z_3$ boson, in contrast, prefers to be in the region of smaller $g_H$ and larger $g_X$.
The mixed composition of $Z_3$ lies in the range of $7\times10^{-4} < g_H < 5\times10^{-3}$ and 
$4\times10^{-4} < g_X < 3\times10^{-3}$. For the $X$-like $Z_3$ boson region in Fig.~\ref{lightMX_mz1_gH_like}, 
there is a lower bound for $g_H$ due to our choice of 200 TeV as the upper bound for $v_\Phi$. Moreover, in Fig.~\ref{lightMX_mz1_gX_like}, one can also see that the $\chi^2$ tolerance sets an upper limit on $g_X$ as the $Z_1$ boson mass gets heavier.
 
Finally, we would like to emphasize that the contact interaction exclusion regions
at $M_{Z_1}<200\mev$ and $10^{-4}<g_X,g_H<10^{-3}$ are owing to two different coupling components,  
$g_H\mathcal{O}_{2i}$ and $g_X\mathcal{O}_{3i}$  
in Eqs.~\eqref{eq:vecfacmy0} and~\eqref{eq:axifacmy0}.

\subsection{Future Prospects\label{subsection:future}}

Since current LEP together with other constraints already 
put a severe limit on the parameter space, it will be interesting to see 
whether the future $Z$-boson precision experiments can further probe our model.  
In the near future, there are three colliders that can
improve $Z$-boson measurements: 
CEPC~\cite{CEPC-SPPCStudyGroup:2015csa}, ILC~\cite{Fujii:2017vwa}, and FCC-ee~\cite{dEnterria:2016fpc}. 
Among them, CEPC is the one that could give the most sensitive limit.    
Therefore, in this subsection, we make an estimation of our parameter space 
with the projected CEPC sensitivity. 

In the third column of Table~\ref{table:ewobs}, we quote the
expected CEPC sensitivity~\cite{CEPC-SPPCStudyGroup:2015csa}. Apparently, some of the error bars are
expected to be significantly reduced.   
Note that the CEPC preliminary conceptual design report does not provide a
full list as the LEP measurements showed in the 2nd
column. Therefore, for those missing rows, 
we reuse the data from the 2nd column (LEP data).

\begin{figure}
	\includegraphics[width=0.49\textwidth]{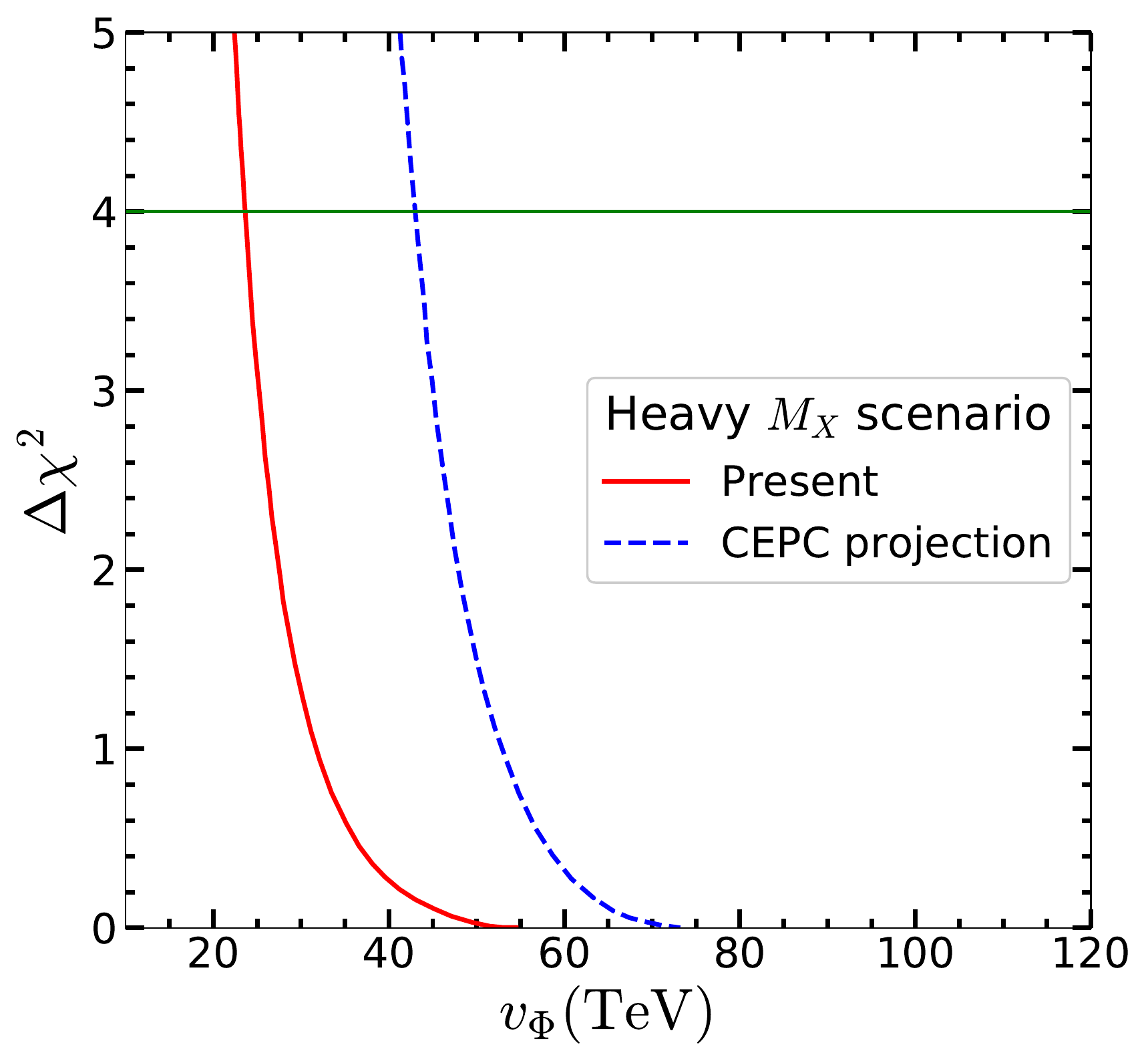}
	\includegraphics[width=0.49\textwidth]{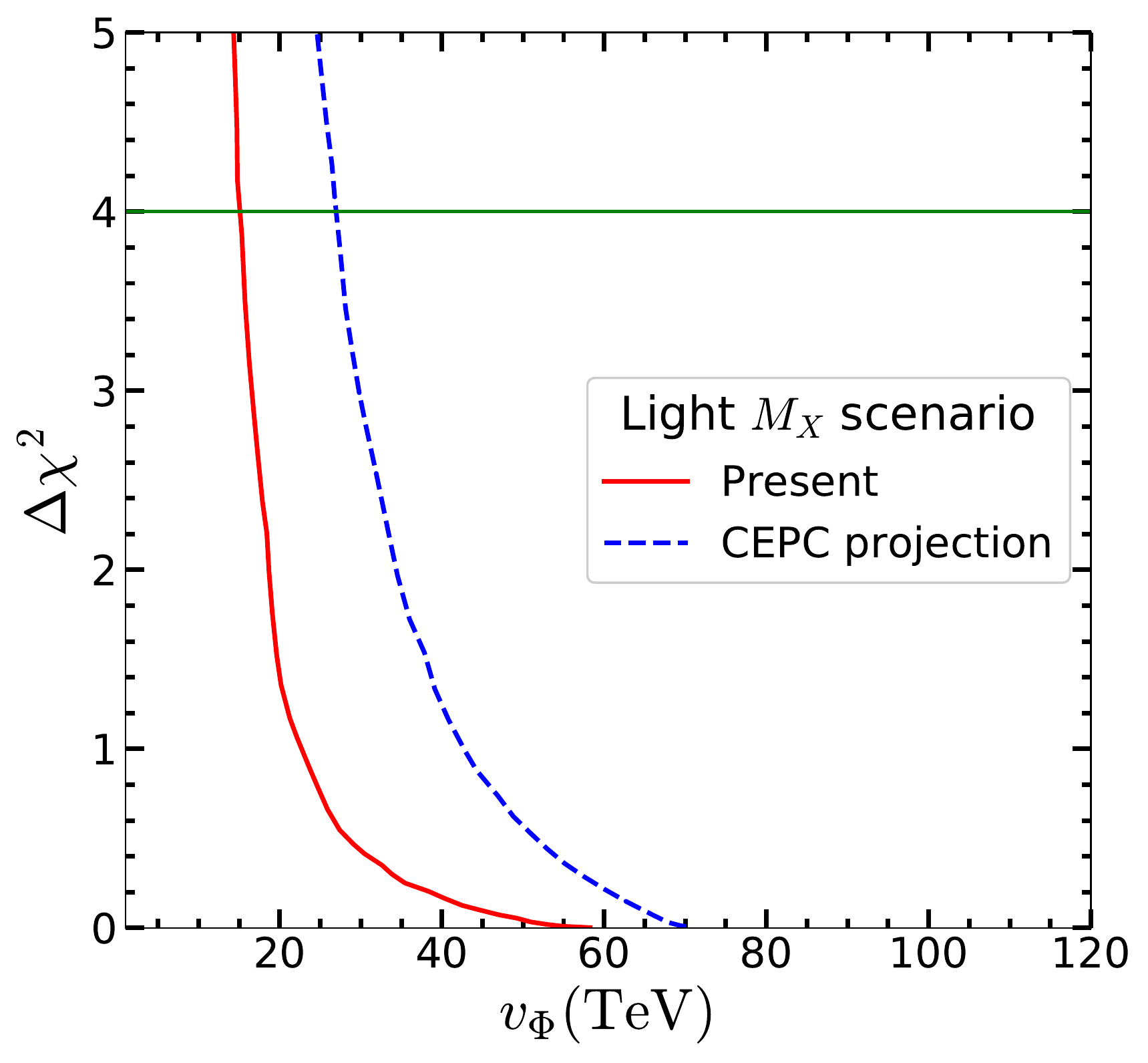}
	\caption{\small The $\Delta\chi^2$ as function of $v_\Phi$.
	The red solid line and blue dashed line are based on present constraint 
	and future CEPC sensitivity. The left and right panels are corresponding to 
	heavy and light $M_X$ scenarios, respectively. 
	} \label{Fig:Fvp}
\end{figure}

To start with, we present the $\Delta\chi^2$ in terms of $v_\Phi$ in
Fig.~\ref{Fig:Fvp} for heavy (left) and light (right) $M_X$ scenarios.
Importantly, $v_\Phi$ is the most sensitive
parameter in the G2HDM, determining the theory scale. For the heavy
$M_X$ scenario, in the present sensitivity case the
$2\sigma$ lower bound is around $24\tev$, while in the CEPC
case it is around 44~TeV. 
For the light $M_X$ scenario, the 2$\sigma$ current and CEPC lower limit of
$v_\Phi$ is smaller than the heavier $M_X$ scenario. In particular, $v_\Phi >
15 \text{~TeV} \,(36\text{~TeV})$ at $95\%$ C.L. from current experiments
(CEPC).
The difference between these two scenarios is owing to the different sources of
constraints on $v_\Phi$.  For the heavy $M_X$ scenario,
the EWPT constraints of the SM $Z$ boson play an important role in
raising the lower limit of $v_\Phi$. However, for the light $M_X$ scenario, the main
constraint to exclude the lower $v_\Phi$ region is from $Z'$ searches. 
This also explains why the future
sensitivity does not further push $v_\Phi$ in the light $M_X$ scenario to larger values as 
the heavy $M_X$ scenario does because 
the future sensitivities of contact interactions are not available for CEPC and only the previous 
limits from LEP II are used.

\begin{figure}
	\begin{flushright}
  	\includegraphics[width=0.49\textwidth]{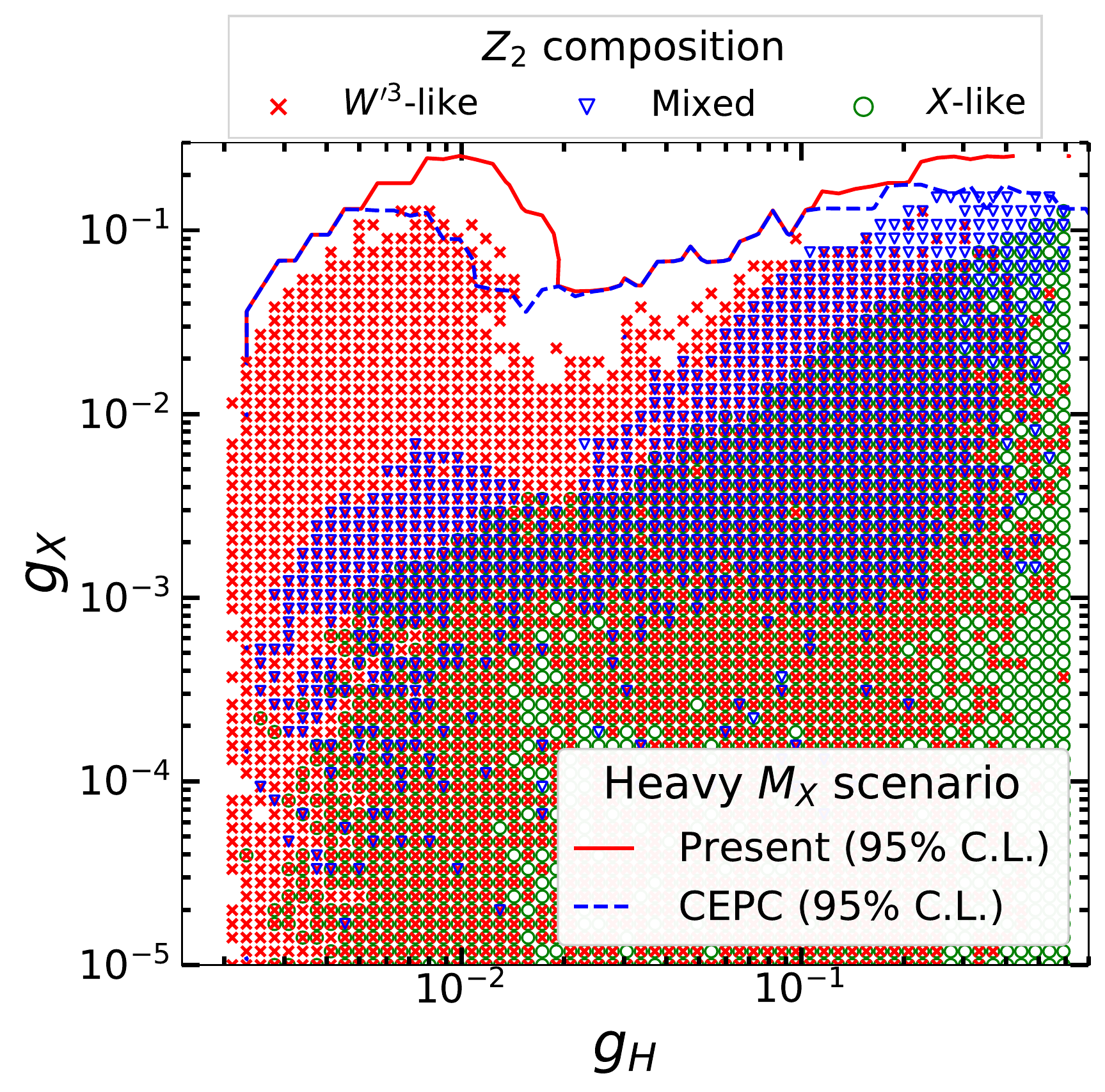}
	\includegraphics[width=0.49\textwidth]{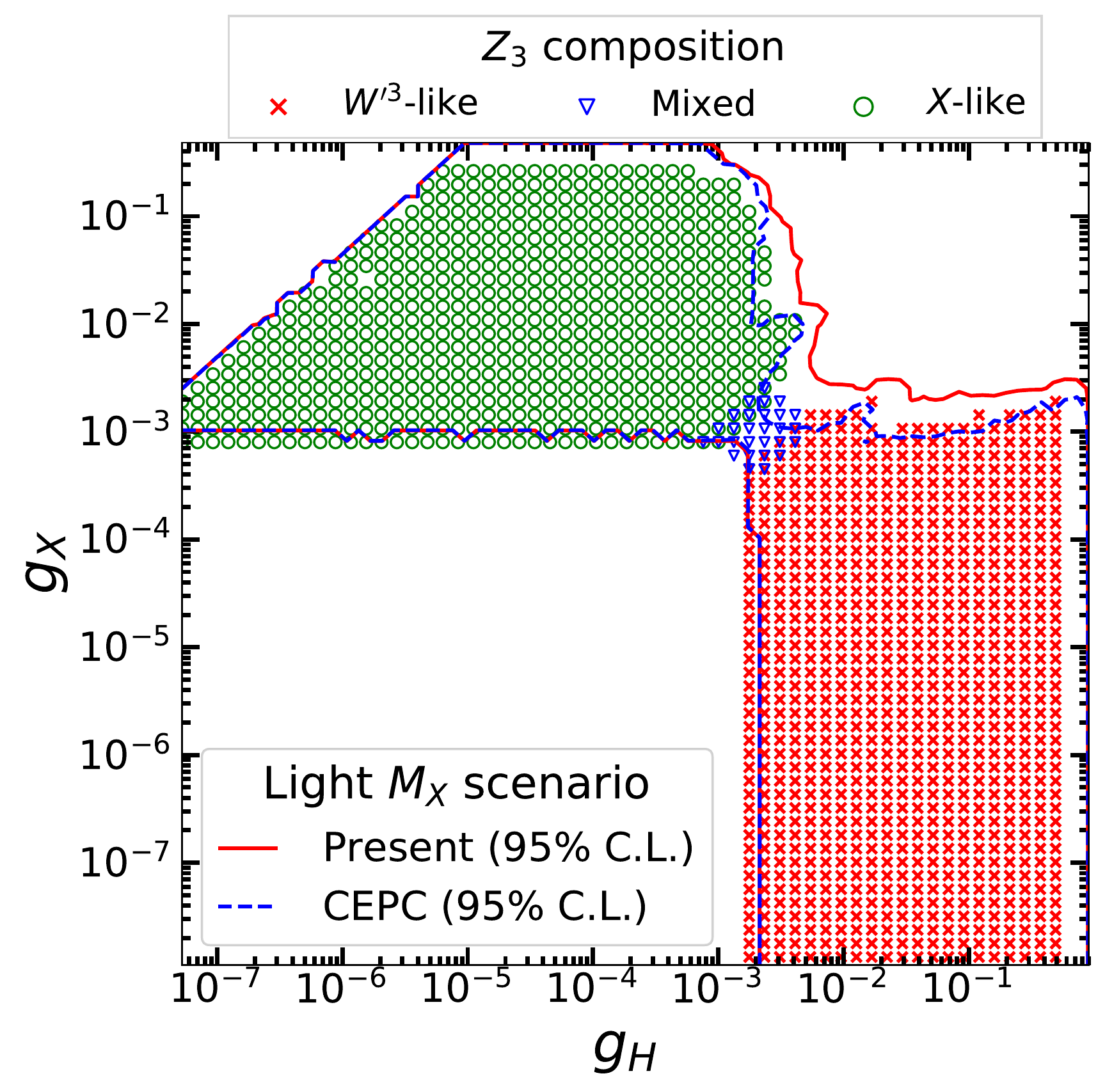}
	\end{flushright}
	\caption{\small 
	The present and future sensitivity allowed regions projected on the ($g_H,~g_X$) plane
	in both heavy (left) and light (right) $M_X$ scenarios.
	The red solid line is for the present $95\%$ limit while the blue dashed line is for the CEPC future sensitivity. 
	Scatter points in $1\sigma$ region of the present constraint are also shown. The color codes in the left and right panel are same as in 
	Figs. \ref{heavyMX_mzp_oi2} and Fig. \ref{lightMX_mzp_oi3}, respectively.
	 	}
	\label{Fig:Fgx}
\end{figure}

In Fig.~\ref{Fig:Fgx}, we compare the present limit and
future CEPC sensitivity of the two-dimensional contours on the ($g_H,~g_X$) plane.
The figure in the left (right) column corresponds to the heavy (light) $M_X$ scenario. 
Because the upper scan limit of $g_H$ is set to be less than $g_2^{\texttt{SM}}$,  
the experimental constraints on $g_H$ are not present. 
In contrast, $g_X$ has an upper limit from the constraints due to $g_H$
having a lower limit from the maximum scanned $v_\Phi$ value. 
The upper limit of $g_H$ can be further improved by future CEPC sensitivity along the edge of the contour. 
However, the light $M_X$ scenario is mildly constrained by future CEPC sensitivity.
The two contour plots in Fig.~\ref{Fig:Fgx} can be further understood as follows. 
We note that, for the case of $W'^3$-like $Z_2$ in heavy $M_X$ scenario (left panel)
or $W'^3$-like $Z_3$ in light $M_X$ scenario (right panel), 
$g_H$ has a lower limit at $\sim 2\times 10^{-3}$ due to our choices of the parameter scan ranges.
Indeed, in both cases we have
$M_{Z_{2,3}} \approx 0.5 g_H \sqrt{v^2+v_\Phi^2}$ which implies that $g_H \approx (2 M_{Z_{2,3}})/\sqrt{v^2+v_\Phi^2}$. 
Since we require $M_{Z_{2,3}} > 210$ GeV and $v_\Phi < 200$ TeV, this implies $g_H > 2 \times 10^{-3} $.
Similarly, for the case of $X$-like $Z_3$ in the light $M_X$ scenario (upper right panel), the mass of $Z_3$, is given by 
$M_{Z_3} \approx \sqrt{g_X^2 (v^2+v_\Phi^2) + M_X^2}$ so that we can obtain $g_X \approx \sqrt{M_{Z_3}^2 - M_X^2}/\sqrt{v^2+v_\Phi^2}$. This yields a lower limit for $g_X$ at $\sim 10^{-3}$ when we require $M_{Z_3} > 210$ GeV, $M_X < 80$ GeV and $v_\Phi < 200$ TeV. On the other hand, $g_X$ has no lower limit in the heavy $M_X$ scenario (left panel).

The St\"{u}eckelberg mass parameter $M_X$ is a filter to split the parameter space into two scenarios 
but we have not been able to constrain this parameter. 
The reason is simply that $Z_3$ in the heavy $M_X$ scenario is too heavy to 
be relevant by current experiments. 
On the other hand, in the light $M_X$ scenario with $M_X<80\gev$, it is again too light 
to be presented in the EWPT data. To constrain light $M_X$, just like dark photon, 
the lightest $Z_1$ could be detected by those future beam dump experiments 
such as NA62~\cite{1510.00172}, Belle II~\cite{1002.5012}, and SHiP~\cite{1504.04855}.
However, this is beyond the scope of this work and we will
return to it in the future. 

\section{Summary and Conclusion} \label{section:conclusions}

In this paper, we perform an updated profile
likelihood analysis for the gauge sector of G2HDM.

For the two St\"{u}eckelberg mass parameters $M_Y$ and $M_X$ associated with the 
hypercharge $U(1)_Y$ and the extra $U(1)_X$ respectively, 
we showed that a nonzero $M_Y$ would produce non-standard QED couplings 
for all the fermions in G2HDM, albeit we can always achieve a massless photon
for arbitrary values of $M_X$ and $M_Y$. 
We therefore set $M_Y = 0$ in our numerical analysis. 
The remaining new parameters in the gauge sector of G2HDM needed to be constrained 
are $g_H$, $g_X$, $v_\Phi$ and $M_X$.

We have examined the remaining parameter
space with the EWPT LEP data at the $Z$-pole, 
contact interaction constraints from LEP-II
and LHC Run II data for the search of high-mass dilepton resonances.  
The contact interactions constraints can definitely provide a lower limit on $v_\Phi$,
but the EWPT data play a significant role to constrain the parameter space
non-trivially. While the LHC search for the high-mass dilepton resonances also impose
important constraints on the parameter space, the Drell-Yan data from the $Z$ decay
does not impose noticeable impacts yet.

We classify our parameter space based on three
different composition ($X$-like, $W'^3$-like, and mixed) of the heavy
neutral gauge boson, either $Z_2$ or $Z_3$, which is the next-heavier $Z$ boson than the SM one, 
in order to manifest the physics and constraints discussed in this paper.   

In the heavy $M_X$ scenario ($M_X>100\gev$), the SM-like $Z$ is the lightest $Z_1$ boson 
and EWPT constraints exclude the small $v_\Phi$ region up to $24\tev$ at $2\sigma$ significance.
However, the EWPT constraints are not so sensitive to the light $M_X$ scenario
($M_X<80\gev$) where SM-like $Z$ is the next-lightest $Z_2$ boson. 
In particular, the $v_\Phi$ is required to be greater than $15\tev$ due to 
the constraints of $Z'$ contact interaction search from LEP-II 
and high-mass dilepton resonance search from LHC Run II. 
Furthermore, in both light and heavy
$M_X$ scenarios, $M_X$ is just a parameter to tweak between two scenarios and
it is totally unbounded in this study. It is likely that the future dark
photon searches might set a limit on the $M_X$ in the light $M_X$ scenario. 
On the other hand, it is not so trivial for the couplings $g_X$ and $g_H$ because we
found it is hard to set an upper bound on them individually.  

Although the SM $Z$ boson is fixed at the $Z$-pole, the allowed physical
masses of the heavier $Z_i$ still depend on the $M_X$ and detailed
composition.  Generally speaking, the $Z_2$ allowed mass range in the heavy
$M_X$ scenario is same as the range of $M_X$ but $Z_3$ mass can reach up to
$70\tev$ for $X$-like composition and $40\tev$ for both $W'^3$-like and mixed
composition.  Like the role of $M_X$ in the heavy scenario, the $M_{Z_1}$ in
the light $M_X$ scenario is dominated  by $M_X$ and the allowed mass ranges of
$M_{Z_1}$ have no difference between different composition.  However,
regarding to $M_{Z_3}$, mixed $Z_3$ is restricted to less than $500\gev$ but
the masses of $X$-like and $W'^3$-like $Z_3$ are below $70\tev$.

Finally, we also discuss the future sensitivity of the new parameters at the CEPC.  
We found that the CEPC can significantly probe the parameter space of the heavy $M_X$ scenario
but the sensitivity is not improved much for the light $M_X$ scenario.   
In the latter case, when $M_X$ is getting very light, 
$Z_1$ can be much lighter than the $Z$-boson and it is more
appropriate to identify it as the dark photon or dark $Z_D$. 
The very rich phenomenology of light dark photon or dark $Z_D$ in G2HDM 
remains to be explored in the future.

\newpage 

\section*{Acknowledgments}
%
We would like to thank Wei-Chih Huang, Zuowei Liu and Xun Xue for stimulating discussions.
This work was supported in part by the Ministry of Science and Technology
(MoST) of Taiwan under Grant Nos. 107-2119-M-001-033- and 107-2811-M-001-027-.
Y.-L.~S. Tsai was funded in part by Chinese Academy of Sciences Taiwan
Young Talent Programme under Grant No. 2018TW2JA0005.
V.~Q. Tran was funded in part by the National Natural Science Foundation of China 
under Grant Nos.\ 11775109, U1738134 
and by the National Recruitment Program for Young Professionals.

	
\appendix

\section{The Rotation Angles $\phi$, $\theta$ and $\psi$}
\label{appendix:euler}

In this appendix, we will show how to obtain the equations of the rotation angles
such as Eqs.~\eqref{eq:Tanphi}, \eqref{eq:Tantheta} and \eqref{eq:Cotpsi} from
the orthogonal matrix which diagonalizes the mass matrix 
$\mathcal{M}^2_{\rm gauge}(M_Y=0)$ given in Eq.~\eqref{eq:MgaugeSMrot}.
The orthogonal matrix we choose is Eq.~\eqref{eq:Rpara} because it
is rather convenient to find all the ${\cal O}_{ij}$s and determine the
rotational angles $\phi$, $\theta$ and $\psi$ numerically. However, the computation of
the angles in terms of the fundamental parameters in the Lagrangian 
are difficult to organize into nice forms using Eq.~\eqref{eq:Rpara} for  $\mathcal{O}$, 
so we apply Cramer's rule for solving the secular equations 
and get another form for  $\mathcal{O}$ as follows
\begin{equation}
\label{Oij}
\mathcal O \ =\   \left( \begin{array}{ccc}
|x_1|/\Delta_1 & x_2/\Delta_2 & x_3/\Delta_3 \\
y_1/\Delta_1 & |y_2|/\Delta_2 & y_3/\Delta_3 \\
z_1/\Delta_1 & z_2/\Delta_2 & |z_3|/\Delta_3 \end{array} \right)\, ,
\end{equation}
where
\begin{equation}
\label{Deltas}
\Delta_i = \sqrt{ x^2_i\, +\, y^2_i\, +\, z^2_i} \, ,
\end{equation}
and 
\begin{eqnarray}
		\label{xyz}
		x_1=\!\left| \begin{array}{lr}
			M^2_{22}-M^2_{Z_1} & M^2_{23} \\
			M^2_{32} & \hspace{-1.2cm} M^2_{33} - M^2_{Z_1} \end{array}\right|\!
		\, ,\
		y_1 = {\rm s}_{x_1}\! \left| \begin{array}{cc}
			M^2_{23} & M^2_{21} \\
			M^2_{33} -M^2_{Z_1} & M^2_{31} \end{array} \right| \, ,\
		z_1 = {\rm s}_{x_1}\! \left| \begin{array}{cc}
			M^2_{21} & M^2_{22} - M^2_{Z_1} \\
			M^2_{31} & M^2_{32} \end{array} \right| \, ,\nonumber\\
		x_2 = {\rm s}_{y_2}\! \left| \begin{array}{cc}
			M^2_{13} & M^2_{12} \\
			M^2_{33} - M^2_{Z_2} & M^2_{32} \end{array} \right| \, ,\
		y_2 = \!\left| \begin{array}{lr}
			M^2_{11} - M^2_{Z_2} & M^2_{13} \\
			M^2_{31}  & \hspace{-1.2cm} M^2_{33} - M^2_{Z_2} \end{array} 
		\right|\!  \, ,\ 
		z_2 = {\rm s}_{y_2}\! \left| \begin{array}{cc}
			M^2_{12} & M^2_{11} - M^2_{Z_2} \\
			M^2_{32} & M^2_{31} \end{array} \right| \, , \nonumber \\
		x_3 = {\rm s}_{z_3}\! \left| \begin{array}{cc}
			M^2_{12} & M^2_{13} \\
			M^2_{22} - M^2_{Z_3} & M^2_{23} \end{array} \right|,\
		y_3 = {\rm s}_{z_3}\! \left| \begin{array}{cc}
			M^2_{13} & M^2_{11} - M^2_{Z_3} \\
			M^2_{23} & M^2_{21} \end{array} \right| \, ,\ 
		z_3 = \!\left| \begin{array}{lr}
			M^2_{11} - M^2_{Z_3} & M^2_{12} \\
			M^2_{21} & \hspace{-1.2cm} M^2_{22} - M^2_{Z_3} \end{array}
		\right|\!  \, ,\nonumber \\ 
\end{eqnarray}
with $M^2_{ij}$ stands for the element of 
$\mathcal{M}^2_{\rm gauge}(M_Y=0)$, $M^2_{Z_i}(i=1,2,3)$ are  
the mass eigenvalues and  ${\rm s}_{x_i} = {\rm sign} (x_i)$. From Eq.~\eqref{eq:Rpara}, 
one can obtain the following relations for the rotational angles 
$\phi$, $\theta$ and $\psi$~\footnote{We note that similar approach had been used in~\cite{Pilaftsis:1999qt}
for the scalar boson mass matrix in MSSM with explicit CP violation.}
\begin{equation}
\label{element in O}
\phi=\arctan{\left(\frac{-\mathcal{O}_{12}} {\mathcal{O}_{22}}\right)} ,\,
\theta=\arctan\left ({\frac{-\mathcal{O}_{32}}{\mathcal{O}_{12}}} \sin{\phi}\right) , \,
\psi={\rm \arccot}\left({\frac{-\mathcal{O}_{21}}{\mathcal{O}_{31}}} \frac{\cos{\theta}}{\sin{\phi}} - \sin{\theta} \cot{\phi}\right) \, ,
\end{equation}
with the range for $\theta$ covers $\pi$ radians, and the range for $\phi$ and $\psi$ covers $2\pi$ radians.
Note that the expressions in Eq.~\eqref{element in O} do not depend on the $\Delta_i$ given in Eq.~\eqref{Deltas}.
Using Eqs.~\eqref{Oij} and \eqref{xyz} for the various $\mathcal O_{ij}$ in Eq.~\eqref{element in O},
after some algebra, one can obtain 
Eqs.~\eqref{eq:Tanphi}, \eqref{eq:Tantheta} and \eqref{eq:Cotpsi}, which are
collected here again for convenience.
\begin{equation}
\label{eqA:Tanphi}
\tan({\phi}) = \frac{- g_H v M_{Z^\text{SM}} (M_X^2 - M_{Z_2}^2 + 2 g_X^2 v_\Phi^2)}
{2 \Big( M_{Z_2}^4 -  \big( M_{Z^\text{SM}}^2 + M_X^2 + (v^2+v_\Phi^2) g_X^2 \big) M_{Z_2}^2 +  M_{Z^\text{SM}}^2 (M_X^2 + g_X^2 v_\Phi^2)\Big)}\, ,
\end{equation} 
\begin{equation}
\label{eqA:Tantheta}
\tan({\theta}) = \frac{-g_X (M_{Z_2}^2 (v^2-v_{\Phi}^2) + M_{Z^\text{SM}}^2 v_{\Phi}^2)}
{v M_{Z^\text{SM}} (M_X^2 - M_{Z_2}^2 + 2 g_X^2 v_\Phi^2)} \sin{\phi} \, ,
\end{equation}
\begin{equation}
\label{eqA:Cotpsi}
\cot({\psi}) =  \frac{g_H(M_{Z_1}^2-M_X^2 - 2 g_X^2 v_\Phi^2)}{g_X ( g_H^2 v_\Phi^2 - 2 M_{Z_1}^2)} \frac{\cos{\theta}}{\sin{\phi}} 
- \sin{\theta} \cot{\phi} \, .
\end{equation} 
Thus one can compute the rotation angles in terms of the fundamental parameters of the model
which can provide some useful insights in the vanishing limits of $g_H$ and $g_X$ as discussed in Section~\eqref{subsection:masses}.

\section{Decay Widths of New Neutral Gauge Bosons}
\label{appendix:decaywidths}

\begin{figure}
    \subfloat[
    ]{
        \includegraphics[width=0.5\textwidth]{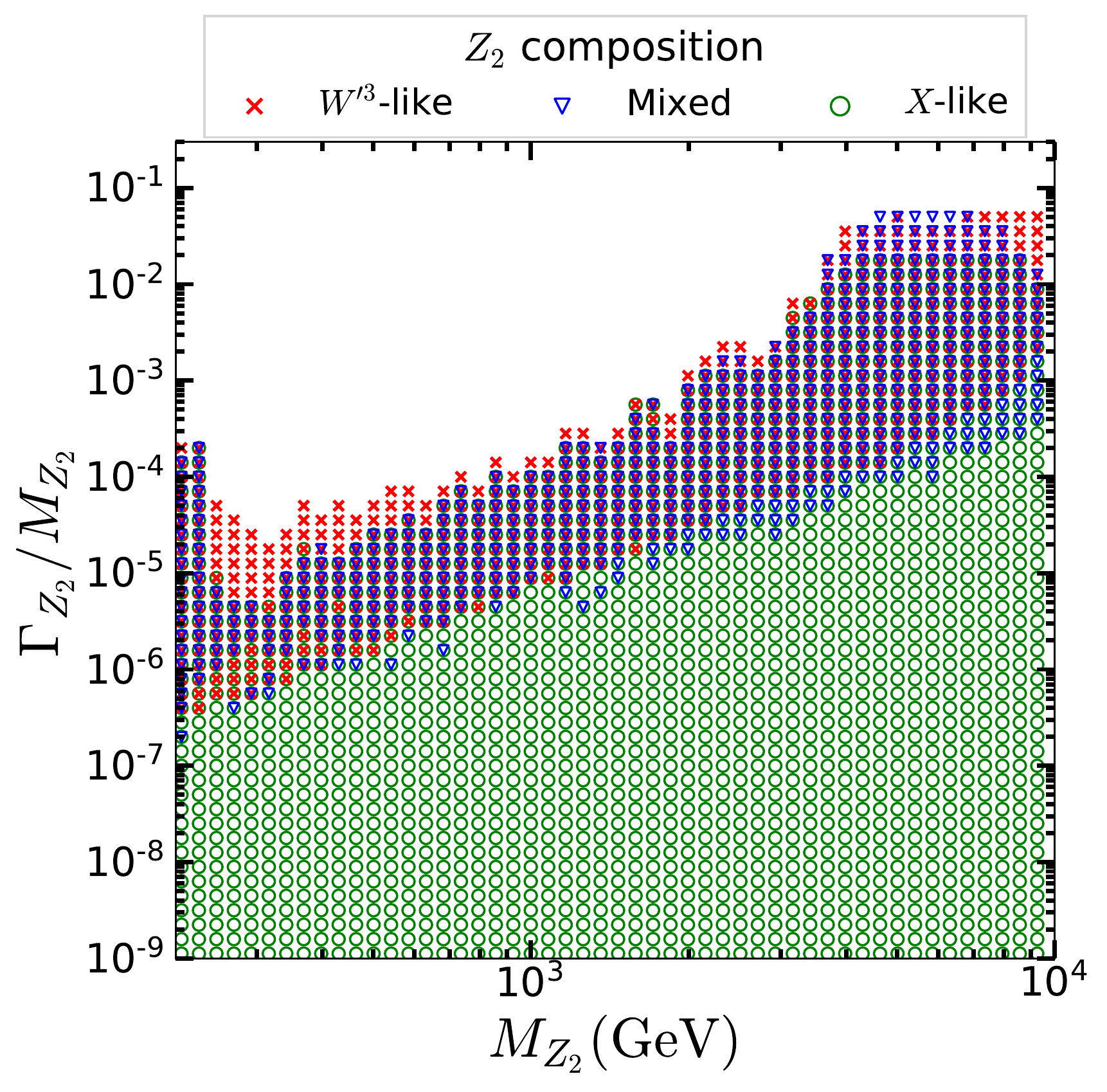}
        \label{heavyMX_zpdecay}
    }
    \subfloat[
    ]{
        \includegraphics[width=0.5\textwidth]{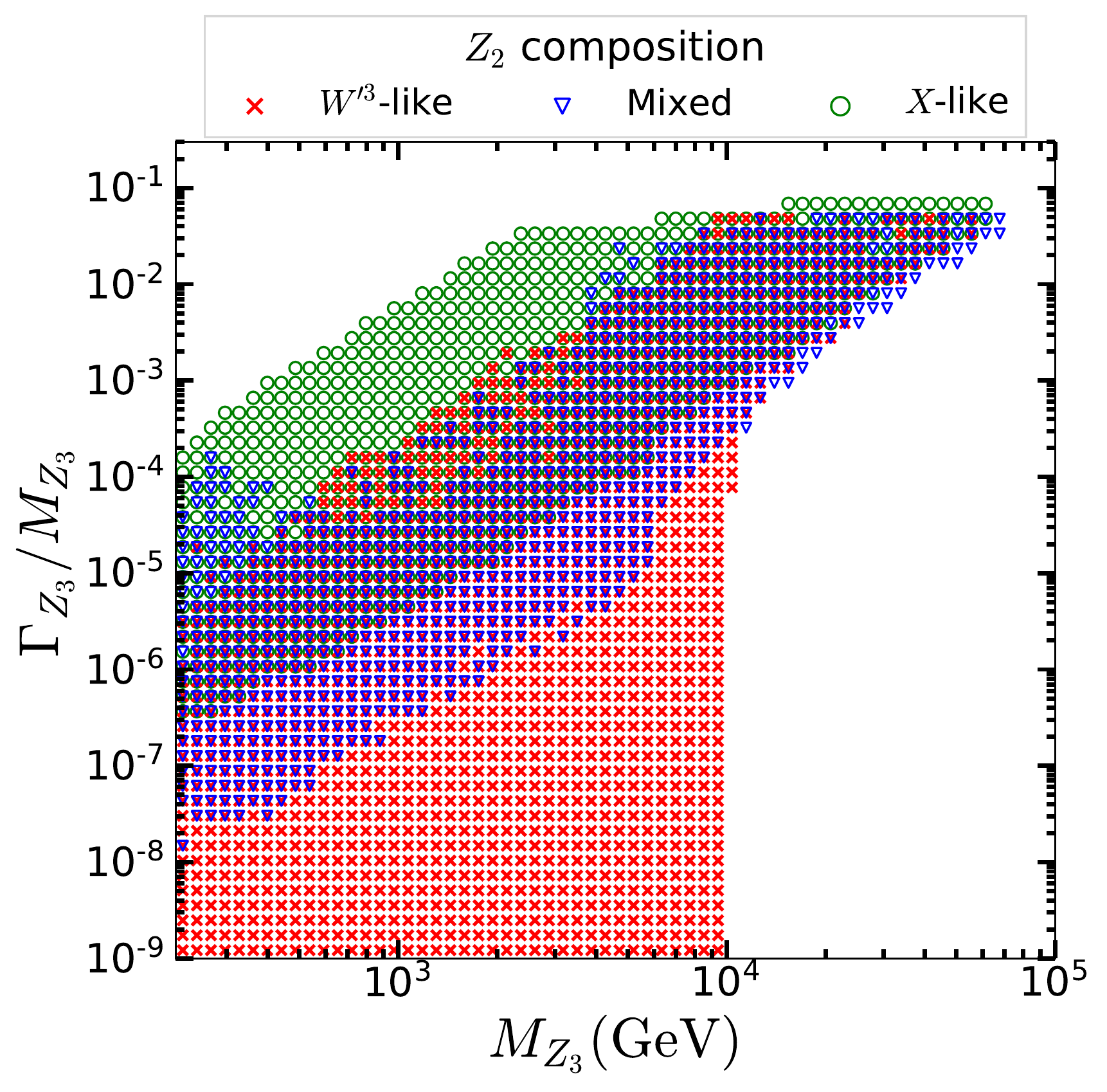}
        \label{heavyMX_zppdecay}      
    }
 \caption{\small Heavy $M_X$ scenario: scatter plots in $1\sigma$ region on the (a) ($\Gamma_{Z_2}$/$M_{Z_2}$, $M_{Z_2}$) 
 and (b) ($\Gamma_{Z_3}$/$M_{Z_3}$, $M_{Z_3}$) planes. 
 The markers are the same as Fig.~\ref{heavyMX_mzp_oi2}. 
 }
 \label{heavyMX_decaywidth}
 \end{figure}

\begin{figure}[h!]
    \subfloat[
    ]{
        \includegraphics[width=0.5\textwidth]{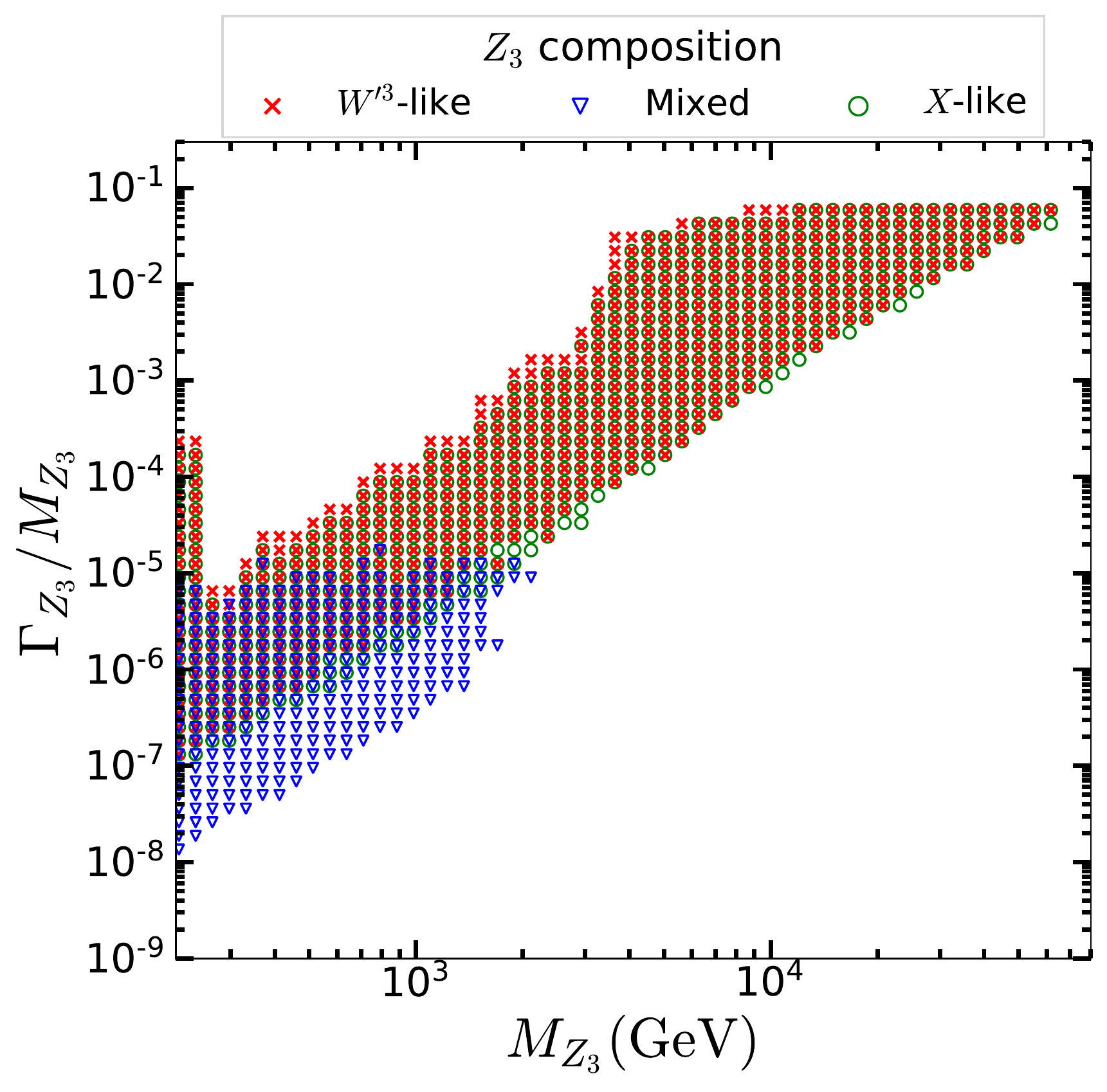}
        \label{lightMX_zpdecay}
    }
    \subfloat[
    ]{
        \includegraphics[width=0.5\textwidth]{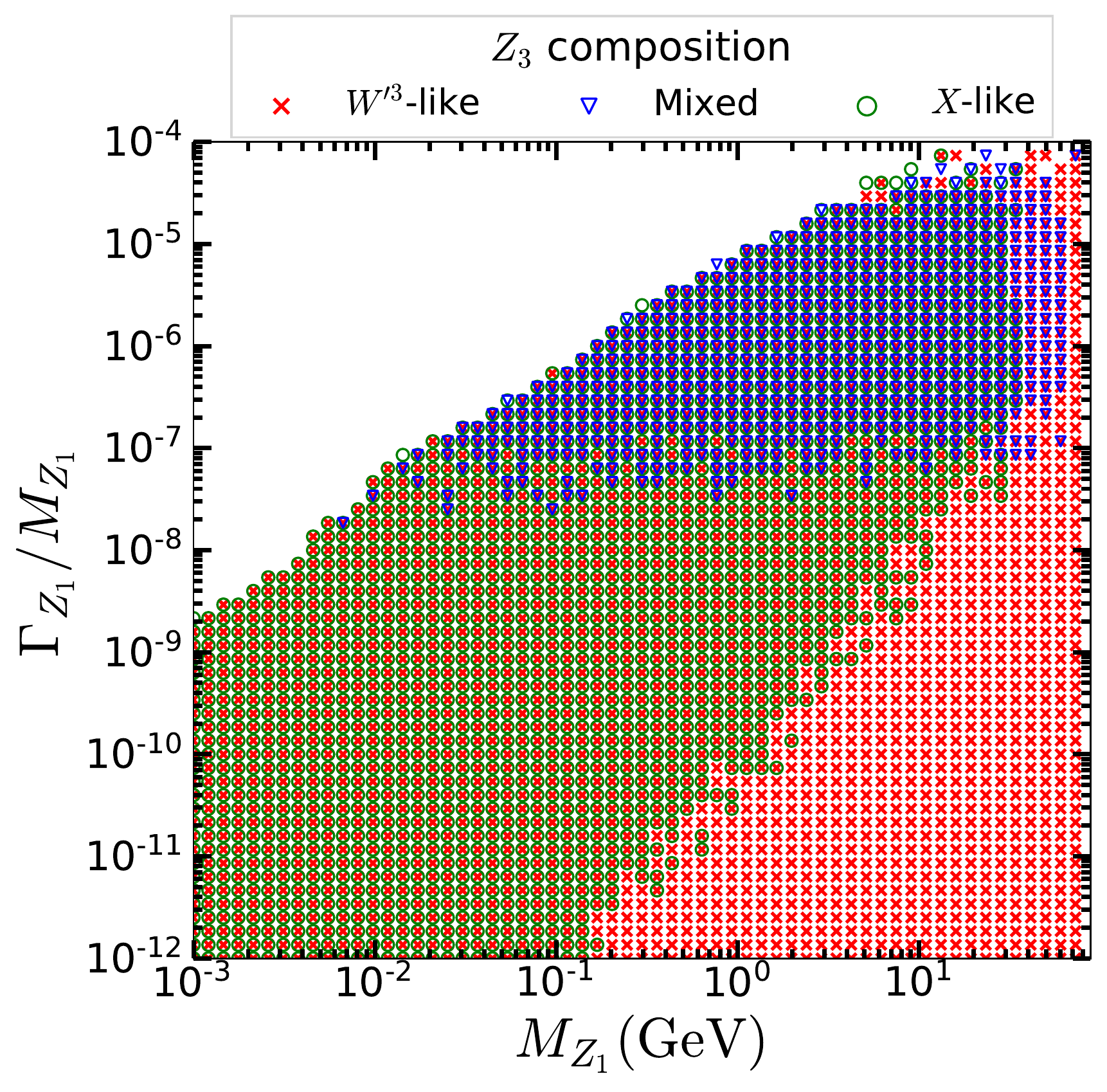}
        \label{lightMX_zppdecay}      
    }
 \caption{\small Light $M_X$ scenario: scatter plots in $1\sigma$ on the (a) ($\Gamma_{Z_3}$/$M_{Z_3}$, $M_{Z_3}$) 
 and (b) ($\Gamma_{Z_1}$/$M_{Z_1}$, $M_{Z_1}$) planes. 
 The markers are the same as Fig.~\ref{lightMX_mzp_oi3}. 
 }
 \label{lightMX_decaywidth}
 \end{figure}

In this appendix, we show the decay widths of the two new neutral gauge bosons $Z_i$. 
We note that for light $M_X$ scenario, $i  = {(1,3)}$, 
while for heavy $M_X$ scenario, $i  = {(2,3)}$. 

\begin{itemize}

\item  
The decay width of $Z_i$ to a pair of fermions (including both SM and new heavy fermions) is given as follows
\beq
\label{eq:Zptoff}
\Gamma(Z_i\to f\bar{f}) =
\frac{N^c_f g_{M}^2 M_{Z_i} }{12 \pi } \sqrt{1-4 r_{if}} \left(  (2 r_{if}+1) |{v^{(i)}_f}|^2 + (1-4 r_{if}) |{a^{(i)}_f}|^2 
\right) \,,
\eeq
where $g_M = \sqrt{g^2+g^{\prime 2}}/2$, $N^c_f$ is the number of color for fermion $f$, 
the coefficients $v_f^{(i)}$ and $a_f^{(i)}$ are the couplings that appear in Eqs.~(\ref{eq:vecfacmy0}) and (\ref{eq:axifacmy0}) and 
$r_{if} = \frac{m_f^2}{M_{Z_i}^2}.$ 

\item
The decay width for $Z_i \to W^+W^-$ process is given by \cite{Barger:1987xw}
\beq
\label{eq:ZptoWpWm}
\Gamma(Z_i\to W^+W^-) =
\frac{g_{Z_iWW}^2 M_{Z_i}}{192 \pi r_{iW}^2  }
(1- 4 r_{iW})^{3/2}
\Big(1 + 20  r_{iW} + 12  r_{iW}^2 \Big) \,,
\eeq
where $r_{iW} = \frac{M_W^2}{M_{Z_i}^2}$  and the coupling $g_{Z_iW W} =  g c_W {\cal O}_{1i}$\,.

\item
Similarly, one can obtain the decay width for $Z_i\to W'^pW'^m$ process as 
\beq
\label{eq:ZptoWppWpm}
\Gamma(Z_i\to W'^pW'^m) =
\frac{g_{Z_iW'W'}^2 M_{Z_i}}{192 \pi r_{iW'}^2  }
(1-4 r_{iW'})^{3/2}
\Big(1 + 20  r_{iW'} + 12  r_{iW'}^2 \Big) \,,
\eeq
where $r_{iW'} = \frac{M_{W'}^2}{M_{Z_i}^2}$  and the coupling $g_{Z_i W' W'} = g_H\,{\cal O}_{2i}$\,.

\item 
The new neutral gauge boson $Z_i$ can also decay into pair of scalar dark matter candidate in this model. 
The decay width for this process $Z_i \to D D^*$ is given by \cite{Barger:1987xw}
\beq
\label{eq:ZptoDD}
\Gamma(Z_i\to D D^*) =
\frac{g_{Z_iDD}^2 M_{Z_i}}{48 \pi  } \,
 (1-4 r_{iD})^{3/2} \,, \eeq 
where the coupling $g_{Z_iDD} = g_H {\cal O}_{2i}$ and  $r_{iD} = \frac{M_{D}^2}{M_{Z_i}^2}$\,. We note that $D$ is a triplet-like scalar dark matter in this model and we assumed this dark matter doesn't mix with other scalars in this calculation. 

\item 
The decay width for $Z_i \to H^+ H^-$ is given by
\beq
\label{eq:ZptoHpHm}
\Gamma(Z_i\to H^+ H^-) =
\frac{g_{Z_iH^+H^-}^2 M_{Z_i}}{48 \pi  } \,
 (1-4 r_{iH^{\pm}})^{3/2} \,,
\eeq
where $r_{iH^{\pm}} = \frac{M_{H^{\pm}}^2}{M_{Z_i}^2}$ and the coupling $g_{Z_iH^+H^-}$ is given as follows
\beq
g_{Z_iH^+H^-} =  \frac{1}{2} (c_W g  - s_W g') {\cal O}_{1i} - \frac{1}{2} g_H {\cal O}_{2i} + g_X {\cal O}_{3i} \,.
\eeq

\item 
The decay width for $Z_i \to Z_j H$ is given by \cite{Barger:1987xw}
\bea
\label{eq:ZptoZH}
\Gamma(Z_i\to Z_jH) = 
\frac{g_{Z_iZ_jH}^2 M_{Z_i}}{192 \pi \,M_{Z_j}^2 } \,
 \Big(1+(r_{ij}-r_{iH})^2 -2 (r_{ij}+r_{iH}) \Big)^{1/2} \nonumber \\
 \times \Big(1+(r_{ij}-r_{iH})^2 +10 r_{ij} - 2r_{iH} \Big) \,,
\eea
where $r_{iH} = \frac{M_{H}^2}{M_{Z_i}^2}$, $r_{ij} = \frac{M_{Z_j}^2}{M_{Z_i}^2}$ and the coupling $g_{Z_iZ_jH}$ is given as follows
\bea
g_{Z_iZ_jH} =  \frac{v}{2} \Big( (c_W g  + s_W g') {\cal O}_{1j} - g_H {\cal O}_{2j} - 2 g_X {\cal O}_{3j} \Big) \nonumber \\
\times \Big( (c_W g  + s_W g') {\cal O}_{1i} - g_H {\cal O}_{2i} - 2 g_X {\cal O}_{3i} \Big) \,,
\eea
here $v$ is the VEV of the SM Higgs field. Note that we have ignored the mixing of SM like-Higgs $H$
with other scalar bosons in the above calculations. 

\item 
Finally, if not kinematically prohibited, the new neutral gauge bosons can also 
decay into $W'$ and the dark matter $D$. The decay width for this process can be computed as
\bea
\label{eq:ZptoDWp}
\Gamma(Z_i\to W^{\prime p} D^* / W^{\prime m} D) = 
\frac{g_{Z_iW'D}^2 M_{Z_i}}{192 \pi \,M^2_{W^\prime} } \,
 \Big(1+(r_{iW'}-r_{iD})^2 -2 (r_{iW'}+r_{iD}) \Big)^{1/2} \nonumber \\
 \times \Big(1+(r_{iW'}-r_{iD})^2 +10 r_{iW'} - 2r_{iD} \Big) \, , \nonumber \\
\eea
where the coupling $g_{Z_iW'D} = g_H^2 {\cal O}_{2i} v_{\Delta}$ 
with $v_{\Delta}$ being the VEV of $SU(2)_H$ triplet Higgs.

\end{itemize} 

In Figs.~\ref{heavyMX_decaywidth} and~\ref{lightMX_decaywidth}, we show the scatter plots of the ratio of decay width over mass 
of the two new gauge bosons in the heavy $M_X$ and light $M_X$ scenarios respectively. 
In those plots, we set the dark matter mass $M_D$ to be $10\%$ of the new heavy neutral gauge boson $Z_i$ ($i.e.$ $M_D = 0.1\times M_{Z_2}$ in the case of heavy $M_X$ scenario, while $M_D = 0.1 \times M_{Z_3}$ in the case of light $M_X$ scenario), 
charged Higgs mass  $M_{H^{\pm}}$ equals 1.5 TeV and the mass of $W^{\prime (p,m)}$ is randomly chosen in the range of [$M_D, 200$ TeV]. 
Moreover, we assume 
the masses of new heavy fermions are degenerate and equal to 3 TeV. 
We note that $v_\Delta$ can be derived from other parameters according to 
$v_\Delta = 0.5 \sqrt{\frac{4 M_{W'}^2}{g_H^2} - (v^2+v_{\Phi}^2)} $. From these scatter plots one can see that 
for the heavy neutral gauge bosons in both scenarios,
their ratios $\Gamma_{Z_i} / M_{Z_i}$ are all below $\sim 1\%$, 
until they are heavier than 10 TeV the ratios can then reach $\sim 6\%$.
However for the light $Z_1$ in the light $M_X$ scenario, $\Gamma_{Z_1} / M_{Z_1}$ is well below $10^{-4}$.

\end{document}